\documentclass[12pt]{article}

\usepackage{scicite}
\usepackage{graphicx}
\usepackage{amsmath}
\usepackage{amssymb}
\usepackage{xcolor}
\usepackage{tabularx}
\usepackage{hyperref}

\usepackage{times}

\newenvironment{sciabstract}{%
\begin{quote} \bf}
{\end{quote}}

\newcommand\lya{Ly$\alpha$}
\newcommand\hi{H\,{\sc i}}
\newcommand\oii{O\,{\sc ii}}
\newcommand\oiii{O\,{\sc iii}}
\newcommand\civ{C\,{\sc iv}}
\newcommand\siv{Si\,{\sc iv}}
\newcommand\hei{He\,{\sc i}}
\newcounter{lastnote}

\title{Extreme damped Lyman-$\alpha$ absorption in young star-forming galaxies at $z=9-11$}

\author
{Kasper~E.~Heintz,$^{1,2\dagger}$ Darach~Watson,$^{1,2}$ Gabriel~Brammer,$^{1,2}$ \\
 Simone~Vejlgaard,$^{1,2}$ Anne~Hutter,$^{1,2}$ Victoria~B.~Strait,$^{1,2}$
 Jorryt Matthee,$^3$ \\ Pascal A. Oesch,$^{4,1,2}$ P\'all Jakobsson,$^5$ Nial R. Tanvir,$^6$ Peter Laursen,$^{1,2}$ \\ Rohan P. Naidu,$^{7,\ddagger}$ Charlotte A. Mason,$^{1,2}$ Meghana Killi,$^{1,2}$ Intae Jung,$^8$ \\ Tiger Yu-Yang Hsiao,$^{9}$ Abdurro’uf,$^{9,8}$ Dan Coe,$^{8,9,10}$ Pablo Arrabal Haro,$^{11}$  \\ Steven L. Finkelstein,$^{12}$ \& Sune Toft$^{1,2}$
\\
\small{$^{1}$Cosmic Dawn Center (DAWN), Denmark} \\
\small{$^{2}$Niels Bohr Institute, University of Copenhagen, Jagtvej 128, DK-2200 Copenhagen N, Denmark} \\
\small{$^{3}$Department of Physics, ETH Zürich, Wolfgang-Pauli-Strasse 27, Zürich, 8093, Switzerland} \\
\small{$^{4}$Observatoire de Genève, Université de Genève, Chemin Pegasi 51, CH-1290 Versoix, Switzerland} \\
\small{$^{5}$Centre for Astrophysics and Cosmology, Science Institute, University of Iceland, Dunhagi 5,} \\ \small{107 Reykjavík, Iceland} \\
\small{$^{6}$School of Physics and Astronomy, University of Leicester, University Road, Leicester, LE1 7RH, UK} \\
\small{$^{7}$MIT Kavli Institute for Astrophysics and Space Research, 77 Massachusetts Avenue,} \\ \small{Cambridge, 02139, Massachusetts, USA}\\
\small{$^{8}$Space Telescope Science Institute (STScI), 3700 San Martin Drive, Baltimore, MD 21218, USA}\\
\small{$^{9}$Center for Astrophysical Sciences, Department of Physics and Astronomy, The Johns} \\ \small{Hopkins University, 3400 N Charles St. Baltimore, MD 21218, USA} \\
\small{$^{10}$Association of Universities for Research in Astronomy (AURA), Inc. for the European} \\ \small{Space Agency (ESA)}\\
\small{$^{11}$NSF's National Optical-Infrared Astronomy Research Laboratory, 950 N. Cherry Ave.,} \\ \small{Tucson, AZ 85719, USA} \\
\small{$^{12}$Department of Astronomy, The University of Texas at Austin, Austin, TX, USA}
\\
\small{$^\dagger$Corresponding author; E-mail: keheintz@nbi.ku.dk} \\
\small{$^\ddagger$NASA Hubble Fellow}
}

\date{}

\begin{document} 


\baselineskip24pt


\maketitle 

\vspace{-1.0cm}


\begin{sciabstract}
% Updated version, focus on single galaxy
The onset of galaxy formation is thought to be initiated by the infall of neutral, pristine gas onto the first protogalactic halos. However, direct constraints on the abundance of neutral atomic hydrogen (\hi) in galaxies have been difficult to obtain at early cosmic times. Here we present spectroscopic observations with JWST of three galaxies at redshifts $z=8.8 - 11.4$, about 400--600\,Myr after the Big Bang, that show strong damped Lyman-$\alpha$ absorption ($N_{\rm H\,\textsc{i}} > 10^{22}$\,cm$^{-2}$) from \hi\ in their local surroundings, an order of magnitude in excess of the Lyman-$\alpha$ absorption caused by the neutral intergalactic medium at these redshifts. 
Consequently, these early galaxies cannot be contributing significantly to reionization, at least at their current evolutionary stages. 
%The \hi\ masses of these large, extended neutral gas reservoirs significantly exceed the galaxies' stellar masses which consequently cannot be contributing to reionization, at least at their current evolutionary stages. 
Simulations of galaxy formation show that such massive gas reservoirs surrounding young galaxies so early in the history of the universe is a signature of galaxy formation in progress. 

%as the galaxies accrete a very large mass of gas from the intergalactic medium as a prelude to rapid star formation. 

%Such large, extended neutral gas reservoirs far exceed the galaxies' stellar masses and are not detected in galaxies at lower redshift, but appear to be common at $z>8$. The abundant local neutral gas reservoirs, high ionization parameters, and spectrophotometric modelling suggest young ($\lesssim 100$\,Myr) stellar populations. Simulations of galaxy formation show that such large, metal-poor neutral gas reservoirs surrounding young galaxies so early in the history of the universe, is a signature of early galaxy formation in progress, as the galaxies accrete a very large mass of gas from the intergalactic medium as a prelude to rapid star formation. Consequently, these early galaxies cannot be contributing to reionization, at least at their current evolutionary stages. 
%These very large damped Lyman-$\alpha$ absorbers are likely to systematically offset the inferred break redshifts -- Is this important for the abstract?
\end{sciabstract}
In the current paradigm of galaxy formation, primordial neutral atomic hydrogen (\hi) accretes onto galaxy halos, cools, and forms stars \cite{Keres05,Schaye10}. Detecting the accretion and build-up of \hi\ gas in the earliest galaxies is therefore central to understanding the first phases of galaxy formation. Constraining the amount of \hi\ in galaxies at $z\gtrsim 8$ is also important to quantify how much ionizing radiation escapes from normal star-forming galaxies at this epoch as they are likely to initiate and serve as the main driver of the large-scale reionization \cite{Stark16}. Unfortunately, direct detection of \hi\ in emission from galaxies at cosmological distances is severely limited by the weakness of the \hi\ 21\,cm hyperfine transition \cite{Fernandez16,Maddox21}.

%Paragraph on potential solution (DLAs) + other high-z H\,\textsc{i}/Lya studies
Alternatively, \hi\ can be measured at high redshifts through the Lyman-$\alpha$ (\lya) absorption feature imposed on the spectra of bright background point sources such as quasars. \lya\ absorbers with column densities $N_{\rm H\,\textsc{i}} \geq 2\times 10^{20}$\,cm$^{-2}$, known as damped \lya\ absorbers (DLAs), reveal the presence of extended, dense gas reservoirs surrounding star-forming galaxies \cite{Wolfe05}. Similar, but far more centrally-located DLAs are also observed within the host galaxies of $\gamma$-ray bursts. However, extended DLAs covering the UV emission of entire galaxies are only observed in the most intensely ionized, metal-poor local galaxies at $z\approx 0$ \cite{McKinney19} or in Lyman-break galaxies at $z\approx 3$ with similar extreme physical properties \cite{Shapley03}, and always with column densities $N_{\rm H\,\textsc{i}} \leq 5\times 10^{21}$\,cm$^{-2}$. With the advent of JWST, we are now able to uncover and characterize in unprecedented detail galaxies during the critical period in the Universe's history, at redshifts $\gtrsim 8$, when the Universe was being reionized and the first galaxies developed. In this paper, we present observations of three early universe galaxies, at redshifts $z\simeq9-11$, which show extreme DLAs with $N_{\rm H\,\textsc{i}} > 10^{22}$\,cm$^{-2}$. We argue that the discovery of this feature in several galaxies of the small sample that has so far been observed suggests that we are now beginning to observe the early development phase for galaxies: the build-up of pristine, neutral gas prior to the production of the bulk of the stars.

% The neutral gas properties have been studied up to redshifts $z\approx 6$ using $\gamma$-ray bursts as probes \cite{Hartoog15,Saccardi23}, however, only in pencil-beam sightlines through their interstellar medium (ISM) and with no additional information about the physical properties of their host galaxies. A crucial link is therefore missing to probe the \hi\ gas reservoirs of typical field Lyman-break galaxies during the epoch of reionization.

%Para presenting search + observations. 
%\paragraph{Observations.} 
We analyse the JWST observations of three select galaxies: two were observed and identified as high-redshift galaxy candidates in JWST/NIRCam imaging obtained through the Cosmic Evolution Early Release Science (CEERS, DD-ERS 1345, PI: S. Finkelstein) Survey \cite{Finkelstein22,Finkelstein23,Bagley23} and later spectroscopically confirmed with dedicated JWST/NIRSpec prism spectroscopic follow-up observations (Prog.~ID: DD-2750, PI: P. Arrabal Haro), CEERS-43833 with MSA ID 28 and CEERS-16943 with MSA ID 1 \cite{ArrabalHaro23}. The third source,
MACS0647-JD, is gravitationally-lensed by the foreground galaxy cluster MACS\,J0647+7015 and was discovered as part of the Cluster Lensing And Supernova survey with Hubble (CLASH) \cite{Coe2013}. JWST/NIRCam and NIRSpec prism observations (observation program GO~1433, PI: D. Coe) show a system with multiple intrinsic components (A and B) which are then multiply-imaged \cite{Hsiao2022}, and the galaxy spectrum covering mainly component A shows a star-forming system at $z=10.17$ based on the identification of seven emission lines \cite{Hsiao23}.

% Data reduction + redshifts
The spectroscopic and imaging data were obtained from the Mikulski Archive for Space Telescopes (MAST) and reduced using {\tt grizli} and a custom-made pipeline (see Supplementary Materials). We scale and match the optimally-extracted 1D spectra to the available JWST/NIRCam photometry using a wavelength-dependent polynomial function. This is to improve the absolute flux calibration and take into account potential slit-losses. 
The reduced and photometrically-calibrated 1D spectra of the three galaxies at $z=9-11$ are shown in Fig.~1. 
We determine the redshift of CEERS-43833 to be $z=8.7622\pm 0.0002$ from the multitude of nebular emission lines detected in the spectrum (see Supplementary Materials, Fig.~S2), consistent with the first analysis of this spectrum presented in \cite{ArrabalHaro23}. Similarly, we determine $z=11.409\pm 0.001$ for CEERS-16943 based on the detection of [\oii]\,$\lambda 3727$, [Ne\,{\sc iii}]\,$\lambda 3870$, and He\,{\sc i}\,$\lambda 3889$. This is the most distant detection of nebular emission lines to date. 
For MACS0647-JD we derive $z=10.170\pm 0.003$, consistent with the previous spectroscopic analysis of this galaxy \cite{Hsiao23}.
To select these galaxies, we began by examining the publicly available JWST spectra for galaxies at $z>8$, looking for any exhibiting a slow rollover at the wavelength region around the \lya\ transition (Lyman-break), indicative of a strong damping wing absorption. Such signatures appear
to be common at these redshifts, although not universal (e.g. \cite{Tang23,Heintz22b}, also see Fig.~S4). From about a dozen candidates, the galaxies selected here offer the three most
robust examples, with clear emission lines pinning down the systemic redshift
and high signal-to-noise (S/N) around \lya. We therefore investigate these three
galaxies as the best exemplars of this particular feature in more detail. 
  
% Describe model and DLA prescription. 
%\paragraph{Analysis.} 
To model the intrinsic continuum galaxy emission, we used the recent set of galaxy templates from \cite{Larson22}, created from BPASS and {\sc Cloudy} models that are specifically designed to reproduce the blue rest-frame ultraviolet (UV) colors of galaxies at $z > 8$. In order to account for absorption from the intergalactic medium (IGM), we model the optical depth of \lya\ due to the Gunn-Peterson effect from an increasingly neutral IGM \cite{Fan06} as
\begin{equation}
    \tau_{\rm GP}(z) = 1.8\times 10^{5} \, h^{-1} \, (\Omega_{\rm DM,0})^{1/2} x_{\rm H\,\textsc{i}} \left( \frac{\Omega_{\rm m,0} \, h^2}{0.02} \right) \left( \frac{1+z}{7} \right)^{3/2} ~ ,
\end{equation}
where $h=H_0 / (100\,{\rm km \, s^{-1} Mpc^{-1}})$ is the present-day dimensionless Hubble parameter, $\Omega_{\rm m,0}$ and $\Omega_{\rm DM,0}$ the current baryonic and dark matter density parameters, and $x_{\rm H\,\textsc{i}}$ is the average neutral to total hydrogen fraction of the IGM.
This equation only holds outside the damping wing of the absorption line, while the damping wing itself will be affected by the velocity distribution of the \hi\ atoms. This correction is given by \cite{MiraldaEscude98} as 
\begin{equation}
        \tau_{\rm IGM}(\lambda_{\rm obs}) = \frac{x_{\rm H\,\textsc{i}} R_\alpha \tau_{\rm GP}(z_{\rm gal})}{\pi} \left( \frac{1+z_{\rm obs}}{1+z_{\rm gal}} \right)^{3/2} 
        \times \left[ I \left( \frac{1+z_{\rm IGM,u}}{1+z_{\rm obs}} \right) - I \left( \frac{1+z_{\rm IGM,l}}{1+z_{\rm obs}} \right) \right]
\end{equation}
following the formalism of \cite{Totani06}. Here, $z_{\rm obs}$ is the observed redshift of the neutral gas, $z_{\rm gal}$ is the galaxy redshift, $R_\alpha = \Lambda_\alpha \lambda_\alpha / (4\pi c) = 2.02\times 10^{-8}$ is a dimensionless quantity that includes the damping constant of the \lya\ resonance $\Lambda_\alpha$, and the \lya\ wavelength $\lambda_\alpha$. This parameterization assumes that the IGM is uniformly distributed within the redshift range $z_{\rm IGM,l}$ to $z_{\rm IGM,u}$. Following \cite{Totani06}, we set the upper bound at the galaxy redshift $z_{\rm IGM,u} = z_{\rm gal}$ and integrate the expression down to $z_{\rm IGM,l} = 6$. The resulting spectral cut-off is not sensitive to the choice of $z_{\rm IGM,l}$ at redshifts much below $z_{\rm gal}$ \cite{Totani06}.
We further add a component describing the optical depth due to \lya\ absorption from the interstellar medium (ISM), or gas in the immediate surroundings of the galaxy, in the shape of a Voigt profile
\begin{equation}
    \tau_{\rm ISM} = C a H(a,x) N_{\rm H\,\textsc{i}}
\end{equation}
where $C$ is the photon absorption constant, $a$ is the damping parameter, and $H(a,x)$ is the Voigt-Hjerting function, following the approximation of \cite{TepperGarcia06}. In the case of strong DLA features, where the Lorentzian wings dominate the shape of the line profile, the optical depth is only sensitive to the column density of neutral atomic hydrogen, $N_{\rm H\,\textsc{i}}$. 

\begin{figure}[!t]
    \centering
    \includegraphics[width=0.8\textwidth]{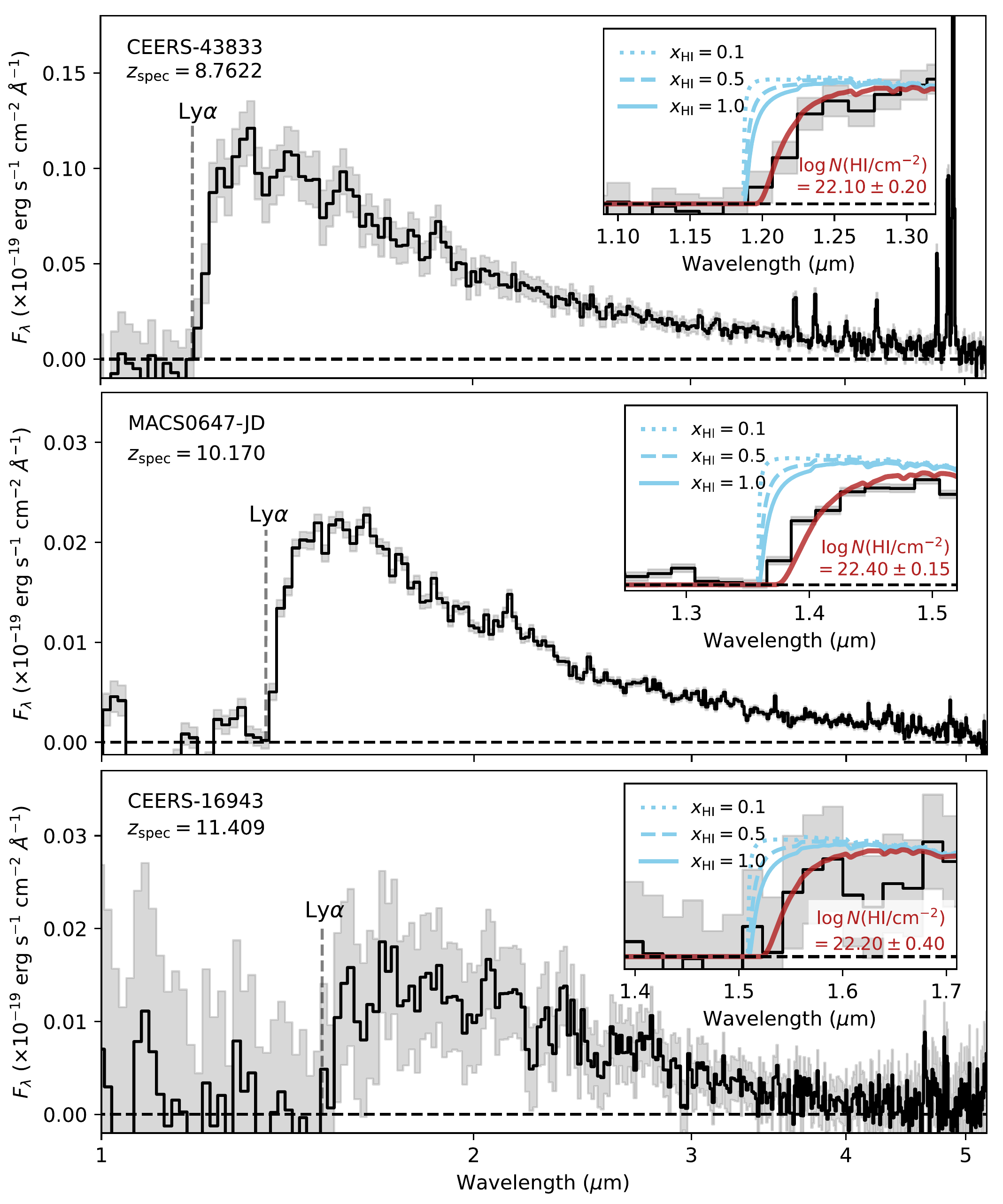}
    \caption{{\bf JWST/NIRSpec spectroscopic data.} Main panels show the reduced and photometrically-calibrated NIRSpec/prism 1D spectra covering $1\mu$m to $5.2\mu$m (black) and the associated $1\sigma$ error spectrum (grey). The galaxy ID and spectroscopic redshifts from the identified emission lines are stated in the top left. The MACS0647-JD spectrum has been demagnified assuming $\mu = 8\pm 1$ \cite{Hsiao23}. 
    In the insets are shown the spectral regions covering the \lya\ transition, with models overplotted showing varying neutral IGM \hi\ fractions, $x_{\rm H\,\textsc{i}}=[0.1,0.5,1.0]$ (blue) and the best-fit model with an additional DLA feature (red). An extremely strong DLA, $\log (N_{\rm H\,\textsc{i}}/{\rm cm^2}) > 22.0$, is detected in all cases.
    }
    \label{fig:fig1}
\end{figure}

% Describe fitting procedure + Fig. 1. 
We model the observed galaxy spectra by fixing the intrinsic galaxy template to the spectroscopic redshifts determined from the detected emission lines, and leave the dust extinction, $A_V$, the neutral \hi\ fraction of the IGM, $x_{\rm H\,\textsc{i}}$, and the column density of the \lya\ absorber from the local galaxy environment, $N_{\rm H\,\textsc{i}}$, as free parameters. Specifically, we use the \texttt{binc100z001age6\_cloudy} model following the recommendation of \cite{Larson22} but note that the exact choice of the intrinsic template, considering for instance older stellar populations, does not have a significant impact on the result ($\lesssim 0.1$\,dex on $N_{\rm H\,\textsc{i}}$). We sample the posterior distributions using the {\tt PyMultiNest}-package \cite{Buchner14}, which uses a Python wrapper around the MultiNest algorithm \cite{Feroz09} to run a multimodal Nested Sampling algorithm on the spectrum. We derive best-fit models with $\log (N_{\rm H\,\textsc{i}}/{\rm cm^{2}})= 22.10\pm 0.20$ for CEERS-43833, $\log (N_{\rm H\,\textsc{i}}/{\rm cm^{2}})= 22.20\pm 0.40$ for CEERS-16943, and $\log (N_{\rm H\,\textsc{i}}/{\rm cm^{2}})= 22.40\pm 0.15$ for MACS0647-JD. The uncertainties represent the 16th and 84th percentiles of the posterior distributions and includes the systematic uncertainty of 0.1\,dex due to variations in the intrinsic galaxy continuum emission templates. 
The models are largely insensitive to the exact value of $x_{\rm H\,\textsc{i}}$, since the line profiles of the DLAs dominate the optical depth at $\lambda_{\alpha}$. To highlight this, we show various models with $x_{\rm H\,\textsc{i}} = 0.1, 0.5, 1.0$ (i.e.\ 10\%, 50\%, and 100\% neutral IGM) in Fig.~1. It is evident that none of these pure IGM models are able to capture the broad absorption features observed in the JWST spectra. While such average sightline models are clearly ruled out even in the case of a fully neutral IGM, we also consider a scenario where these extreme DLAs could be caused by the galaxies residing in overdensities of a largely neutral IGM. However, this seems unlikely since the overdensities required at this epoch would be a factor of $30-50$ larger than average, approaching galaxy ISM densities, a very rare occurrence. We do note, however, that there appears to be an overdensity in the CEERS field at $z\sim 8.7$ \cite{Leonova22,Whitler23}, at least $\times 5$ over the field average for very bright, $M_{\rm UV} < -20$ galaxies, though still insufficient to explain the observed properties. 

We show in the Supplementary Materials, Fig.~S6, the distributions derived from a cosmological hydrodynamical simulation using sightlines from galaxies with similar stellar masses and star-formation rates (SFRs) at $z\approx 8.8$ to investigate the impact of density variability in the IGM. In those simulations \cite{Laursen19}, the 95\% confidence limits are significantly offset from the observed damping wing, albeit assuming a low $x_{\rm H\,\textsc{i}}$ fraction at this epoch. These models thus provide supporting evidence that the most likely interpretation of the DLAs is related to \hi\ gas in the immediate surroundings of the galaxies. We also exclude scenarios involving nebular continuum emission and $2\gamma$ processes as the cause of the damped \lya\ profiles in the Supplementary Materials.
We further compare the CEERS-43833 spectrum to a set of spectra of four galaxies at $z\approx 9$ with similar quality data, i.e.\ high S/N around \lya\ and with robust redshift measurements from nebular emission (Fig.~S4). These comparison galaxies do not show similar broad absorption features, though are consistent with substantial IGM neutral fractions in the line of sight, substantiating the distinct detection of the extreme DLAs in the three $z=9-11$ galaxies in excess of \lya\ absorption from the IGM only. In Fig.~S4, we also compare the observed spectra to the intrinsic IGM and DLA models convolved with the JWST/NIRSpec prism wavelength-dependent spectral resolution, excluding the coarse spectral resolution as the cause of the observed broad \lya\ line profiles.
%\noindent {\bf Fig. 1.}

% Add line ratios, SED fitting, etc. here
To examine the line emission and physical properties of the three $z=9-11$ galaxies in more detail, we measure the line fluxes of each transition detected in the spectra. For CEERS-43833, we detect at high significance the strong nebular emission lines [\oii]\,$\lambda 3727$, [\oiii]\,$\lambda\lambda 4960,5008$, \hei\,$\lambda 3889$ and the Balmer lines H$\beta$, H$\gamma$, and H$\delta$ and recover the same emission lines features previously reported in MACS0647-JD \cite{Hsiao23}. For CEERS-16943 we marginally detect [\oii]\,$\lambda 3727$, [Ne\,{\sc iii}]\,$\lambda 3870$, and He\,{\sc i}\,$\lambda 3889$ at $\approx 3\sigma$ statistical significance. To constrain the line fluxes of each transition, we fit a set of superimposed Gaussian line profiles on the best-fit continuum galaxy model, tying the redshift and line widths of the emission lines, but keep the line fluxes for each transition as free parameters. We report the measured line fluxes in Table~S1. We also detect absorption features from the \siv\,$\lambda 1397$ and \civ\,$1550$ transitions in CEERS-43833 at the same redshift as the nebular emission lines, further hinting at the large, abundant gas reservoir in the local environment of this galaxy. We derive SFRs in the range $1-15\,M_\odot$\,yr$^{-1}$ based on the H$\beta$ or [\oii]\,$\lambda 3727$ line flux measurements and infer relatively low metallicities of $12+\log({\rm O/H}) = 7.4-7.7$, i.e.\ 5--10\% solar abundance values, based on strong-line calibrations (Supplementary Materials). The three galaxies are all characterized by young stellar populations ($\lesssim 100$\,Myr) and have stellar masses in the range $M_\star = 10^{8}-10^{9}\,M_\odot$ based on spectrophotometric fitting of their spectral energy distributions (SEDs), as summarized in Table~1. Overall, these three galaxies have physical properties consistent with the typical star-forming galaxy population at $z=7-10$ \cite{Heintz22b}, see Fig.~S3.

\begin{table}[!t]
    \centering
    \caption{{\bf Physical properties of the three star-forming at $z=9-11$ with strong DLAs.}}
    \begin{tabular}{lccc}
     \hline
     & CEERS-43833 & MACS0647-JD & CEERS-16943 \\
    \hline
    R.A. & 14h19m45.27s & 06h47m53.12s & 14h19m46.36s \\
        \vspace{0.2cm} %% Line for SED
    Decl. & 52d54m42.3s & 70d14m22.98s & 52d56m32.8s \\
    $z_{\rm spec}$ &  $8.7622\pm 0.0004$  & $10.170\pm 0.003$ & $11.409\pm 0.001$ \\
    \vspace{0.2cm}
    $\log N$(\hi/cm$^{-2}$) &  $22.10\pm 0.20$ & $22.40\pm 0.15$ & $22.20\pm 0.40$ \\
    12+log(O/H) & $7.46\pm 0.09$ (O3) & $7.40\pm 0.30$ (Ne3O2) & $7.70\pm 0.30$ (Ne3O2) \\
    SFR$_{\rm H\beta / [OII]}$ / ($M_\odot$\,yr$^{-1}$) & $14.5^{+14.4}_{-7.2}$ & $(1.4\pm 0.2)^a$ & $4.3^{+4.2}_{-2.1}$ \\
    \vspace{0.2cm} %% Line for SED
    $O_{32}$ & $11.4\pm 1.7$ & $(30\pm 6)^a$ & -- \\ 
%    $A_V$ / mag & $<0.15$ \\
    $\log (M_\star / M_\odot)$ & $8.72\pm 0.03$ & $8.1\pm 0.3^a$ & $8.90\pm 0.08$ \\
    SFR$_{\rm SED}$ / ($M_\odot$\,yr$^{-1}$) & $2.3^{+0.2}_{-0.2}$ & $8\pm 3^a$ & $4.8^{+0.7}_{-0.8}$ \\
    $\log U$ & $-1.91^{+0.10}_{-0.15}$ & $-1.9\pm 0.2^a$ & $-1.99^{+0.62}_{-0.64}$ \\
    $A_{\rm V, SED}$ / mag & $0.04\pm0.02$ & $0.07\pm 0.06^a$ & $0.04^{+0.03}_{-0.02}$ \\
    \hline
    \end{tabular}
%    \begin{tablenotes}
    {$^a$Measurements adopted from Hsiao et al.\ \cite{Hsiao23}.}
%    \end{tablenotes}
    \label{tab:galprops}
\end{table}

% Comparisons
Our observations further enable a comparison between the properties of these three young, high-$z$ galaxies to green pea galaxies at redshifts $z\approx 0$, which are considered to be lower redshift analogs to high-$z$ galaxies \cite{McKinney19}. In Fig.~2, we show the \hi\ column density of these extreme $z\approx 0$ emission-line galaxies versus their gas-phase metallicity and ionization parameter $O_{32}$, including also the most metal-poor local galaxies known: I\,Zw\,18, SBS\,$0335-052$, and SBS\,$1415+437$ \cite{Hernandez20}. The three galaxies at $z=9-11$ show substantially higher \hi\ gas columns than any of these local extreme galaxies, even at similar low gas-phase metallicities.
%which may be a selection bias in favour of less dusty systems. 
We also observe a hint of increasing \hi\ column densities for galaxies with higher ionization parameters. This is somewhat at odds with predictions and previous empirical results \cite{Yang17,Jaskot19} that show that larger \lya\ emission equivalent widths and escape fractions are expected from galaxies with higher $O_{32}$ ratios. The complete absorption of \lya\ in these three high-$z$ galaxies might thus be additional signatures of their young stellar populations, which have not yet had time to significantly ionize even their local surroundings, as a prelude to initiating large-scale IGM reionization. Alternatively, the strong DLA features in these particular galaxies could be related to viewing angle effects, indicating that we observe particular gas-rich sightlines, whereas other low column density channels (as indicated by the high $O_{32}$) might exist. Such strongly directionally-dependent escape of ionizing radiation has been proposed by simulations \cite{Cen15,Rosdahl22,Yeh23}. However, due to the complete absorption of \lya, the covering fraction must be larger than the entire rest-frame UV emission of the galaxies in the line of sight. 

\begin{figure}[!t]
    \centering
    \includegraphics[width=0.95\textwidth]{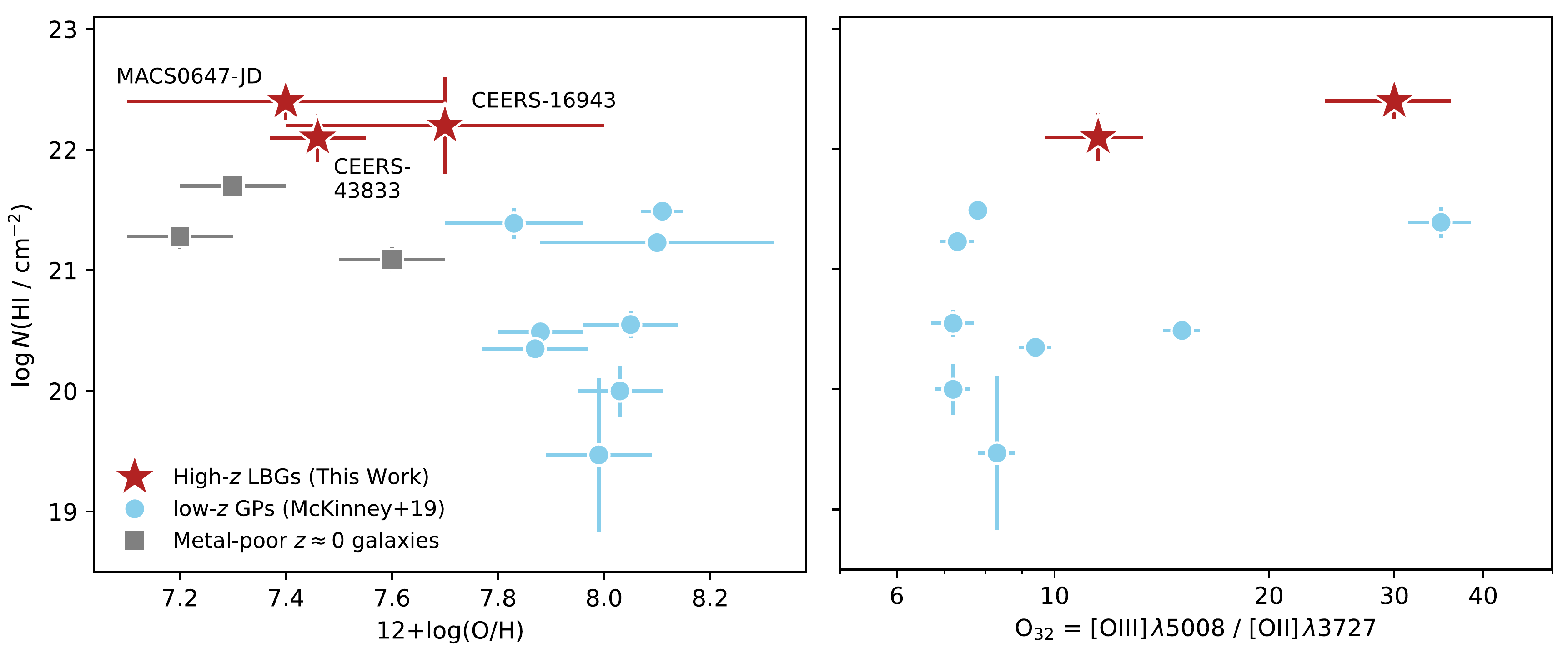}
    \caption{{\bf \hi\ column density relations.} ({\it Left:}) The \hi\ abundances of the three young, $z=9-11$ star-forming galaxies (red stars) compared to local green pea (GP) galaxies (blue dots) \cite{McKinney19} and the three most metal-poor galaxies at $z\approx 0$ (grey squares) \cite{Hernandez20} as a function of gas-phase metallicity, $12+\log({\rm O/H})$. ({\it Right:}) \hi\ abundances as a function of the $O_{32}$ ionization. The symbols denote the same galaxy samples. The $z=9-11$ young star-forming galaxies show the most abundant DLAs.}
    \label{fig:fig2}
\end{figure}

% Implications for massive H\,\textsc{i} gas reservoirs
Based on the large \hi\ column densities and the complete covering fraction of \lya, we can place lower bounds on the \hi\ gas masses of the three $z=9-11$ galaxies. For CEERS-43833 and CEERS-16943, we derive effective areas of the projected ellipses from the measured FWHM major and minor axes of the galaxies, yielding $A_e = 0.67\,{\rm kpc}^{2}$ and $A_e = 0.28\,{\rm kpc}^{2}$, respectively. For MACS0647-JD, we take the distance from the edges of the A and B components to be the effective diameter of the galaxy, $\sim0.5$\,kpc \cite{Hsiao2022}, resulting in an area of $A_e = 0.25\,{\rm kpc}^{2}$. This results in columnar \hi\ gas masses, $M_{\rm H\,\textsc{i}} =  m_{\rm H\,\textsc{i}}N_{\rm H\,\textsc{i}}A_e$, of approx.\ $6.7\times10^{7}\,M_\odot$, $3.5\times10^{7}\,M_\odot$, and $5.0\times10^{7}\,M_\odot$ for the foreground DLAs in the line of sight. 
For a spherical geometry and with gas covering sizes $\times 2-3$ greater than the visible rest-frame UV galaxy radii \cite{Fudamoto22}, this implies total \hi\ gas masses of approx $M_{\rm H\,\textsc{i}}\simeq 10^{8.5} - 10^{9}\,M_\odot$. These \hi\ masses are approx. $\times 2-7$ larger than the inferred stellar masses. Further, they imply short gas depletion times of 1--50\,Myr due to their current active star formation.
The inferred column densities also correspond to \hi\ gas surface densities of $\log (\Sigma_{\rm gas} / M_\odot\,{\rm pc}^{-2}) = 2.0$ (CEERS-43833), $\log (\Sigma_{\rm gas} / M_\odot\,{\rm pc}^{-2}) = 2.1$ (CEERS-16943), and $\log (\Sigma_{\rm gas} / M_\odot\,{\rm pc}^{-2}) = 2.3$ (MACS0647-JD), respectively, larger than the gas surface densities observed in typical galaxies at $z\approx 0$ \cite{Kennicutt12}. 

Based on the SFRs and physical sizes of the galaxies, we derive SFR surface densities of $\log (\Sigma_{\rm SFR} / M_\odot\,{\rm yr^{-1}\,kpc^{-2}}) = 1.3\pm 0.3$ (CEERS-43833) and $\log (\Sigma_{\rm SFR}/M_\odot\,{\rm yr^{-1}\,kpc^{-2}}) = 0.9\pm 0.3$ (CEERS-16943), and approximately $\log (\Sigma_{\rm SFR} / M_\odot\,{\rm yr^{-1}\,kpc^{-2}}) \sim 2$ \cite{Hsiao23} for MACS0647-JD. These measurements place the young, $z=9-11$ star-forming galaxies among the population with the highest SFR surface densities observed in the local Universe \cite{Kennicutt12}. However, their associated gas surface densities place them offset to the Kennicutt-Schmidt (KS) relation, $\Sigma_{\rm SFR} \propto \Sigma^n_{\rm gas}$ with $n\approx1.4$. To be consistent with the KS relation, the gas surface densities should approach $\log (\Sigma_{\rm gas} / M_\odot\,{\rm pc}^{-2}) \sim 3-4$, more than an order of magnitude larger than inferred from the DLA fit. This suggests a potentially large inherent gas mass in the galaxy itself, which does not contribute strongly to the DLA feature. This is similar to what is seen in I\,Zw\,18, where the 21\,cm inferred gas surface density \cite{Lelli2012} is nearly an order of magnitude higher than the DLA-inferred column density \cite{Hernandez20}. Hence, the gas masses roughly estimated above for the \hi\ gas in a shell around the galaxies must be considered as lower limits to the galaxies total gas masses, which may well be greater by an order of magnitude and thus more consistent with the fundamental KS relation. 

\begin{figure}[!ht]
    \centering
    \includegraphics[width=0.7\textwidth]{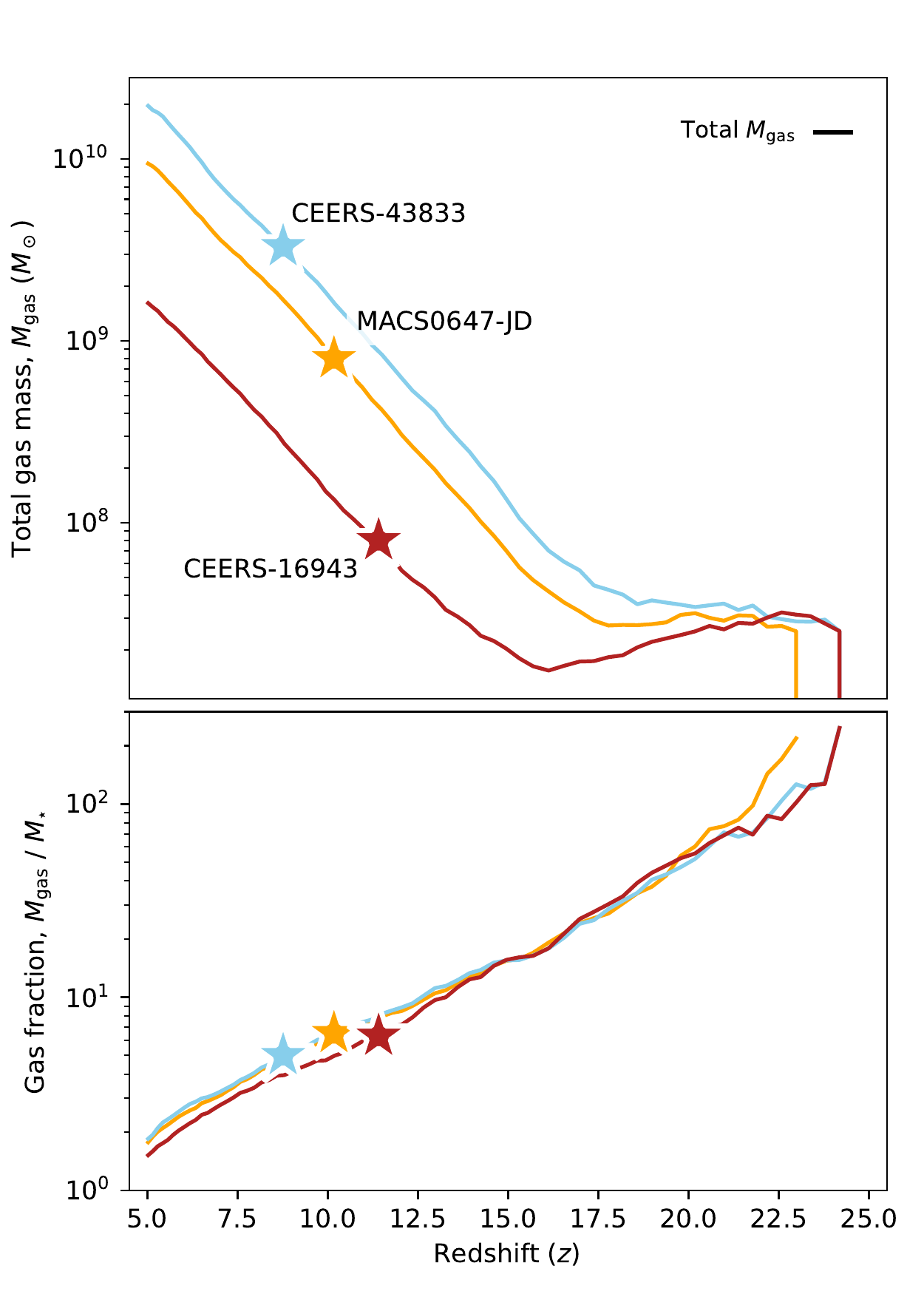}
    \caption{{\bf Gas build-up and accretion histories.} ({\it Top:}) The predicted evolutionary tracks from the Astraeus simulations \cite{Hutter21} of the build-up of $M_{\rm gas}$ in the three $z=9-11$ galaxies based on their stellar masses as a function of redshifts. ({\it Bottom:}) The predicted tracks for the gas fraction, $M_{\rm gas}/M_\star$, for the three $z=9-11$ galaxies. The curves are generally consistent and suggest gas fractions $M_{\rm gas}/M_\star \approx 5$ for the high-$z$ galaxies at their respective redshifts.}
    \label{fig:fig3}
\end{figure}

% Astraeus sims...
To quantify this and make predictions for the overall gas abundances and accretion histories of these three young, $z=9-11$ star-forming galaxies, we extract the mean gas masses of galaxies at their derived stellar masses and redshifts from the {\tt Astraues} simulations framework \cite{Hutter21} (see Supplementary Materials for details). 
The gas mass build-up for this subset of galaxies from $z=25$ to $z=5$, i.e. 130\,Myr to 1.2 Gyr after the Big Bang, in addition to the gas mass accretion at a given redshift, is shown in Fig.~3. From this simulation, we predict gas masses of $M_{\rm gas} = 3\times 10^{9}\,M_\odot$ (CEERS-43833), $M_{\rm gas} = 8\times 10^{7}\,M_\odot$ (CEERS-16943), and $M_{\rm gas} = 8\times 10^{8}\,M_\odot$ (MACS0647-JD). We further infer gas fractions $M_{\rm gas}/M_\star \approx 5$, or $f_{\rm gas} = M_{\rm gas}/(M_{\rm gas} + M_{\star}) \approx 0.85$ for all three galaxies, consistent with the lower bound obtained from the line-of-sight \hi\ column densities. The total gas masses and gas fractions predicted in this simulation overall show gradual increases as a function of decreasing redshift, from $M_{\rm gas} \sim 2\times 10^{7}$ ($M_{\rm gas}/M_\star = 100$) at $z>20$ to $M_{\rm gas} = 10^{9}-10^{10}$ ($M_{\rm gas}/M_\star = 1-2$) at $z=5$. 
%For CEERS-43833 and MACS0647-JD the accreted gas mass appears to be sub-dominant to the {\it in situ} gas mass at all redshifts. Intriguingly, CEERS-16943 shows a larger mass of new, infalling gas than the {\it in situ} gas mass at each point between $z=17$ to $z=10$, with the intersection being close to the observed redshift of the galaxy. This indicates that we are seeing this galaxy as it is still actively accreting new, pristine gas from the IGM. 
%Similarly, the gas fraction starts at $>99\%$ at $z=23$, corresponding to $M_{\rm gas}/M_\star \approx 100$, and decreases to $f_{\rm gas} \approx 85\%$ ($M_{\rm gas}/M_\star \approx 5$) by $z=9$. 

% Discussion!

% Summary paragraph
Our results provide the first direct detection of the abundant neutral atomic \hi\ gas reservoirs in the local environment of galaxies at the onset of the epoch of reionization at $z>8$. Circumstantial evidence for these abundant, neutral gas reservoirs in early galaxies has also been presented in recent literature based on strong absorption from singly-ionized UV absorption features \cite{Boyett23} or an apparent flux deficit from the expected damped \lya\ profile from a neutral IGM \cite{Hsiao23}. Such abundant \hi\ gas reservoirs in early galaxies are also observed indirectly through far-infrared gas tracers \cite{Heintz22a,Heintz23}. Our results imply that extreme \hi\ gas masses might be common in galaxies at $z=9-11$, in stark contrast to the scarcity of similar strong \lya\ absorption features in Lyman-break galaxies at $z\approx 3$ \cite{Shapley03}. 
%High \hi\ column densities and higher metallicities, resulting in large dust columns, may, however, likely bias the observations against the most \hi-rich LBGs at lower redshifts. 
Based on the large abundances of neutral, pristine gas and the fact that these galaxies are surrounded by massive shells or extended sheets of neutral gas we argue that these DLA galaxy systems may be unique to the reionization epoch. Consequently, galaxies like these cannot be leaking significant fractions of ionizing photons due to their extensive gas covering fractions, at least at their current evolutionary stages. These findings have important implications for understanding the process of reionization, and future studies of the incidence and demographics of DLAs surrounding high-redshift galaxies will help distinguish between late and rapid reionization models \cite{Naidu20,Lu23}, over a more smooth and early transition \cite{Finkelstein19}. The discovery of these extreme DLAs during reionization also has important implications for how robustly the degree of ionization of the IGM on large scales can be inferred. 
%Establishing the incidence of DLAs will be important for robust IGM inferences, and may, if not properly taken into account, lead to 
Clearly the damping wings caused by the neutral gas in the IGM will be masked by the large \hi\ abundance in the local environment of galaxies at this epoch, leading to potential overestimations of the degree of neutrality in the bulk of the IGM \cite{McQuinn08} and overall photometric distances of galaxies in the early Universe.
The young stellar populations and strong ionization parameters in the galaxies studied here, in addition to the potential more numerous population of \hi-rich star-forming galaxies at $z>8$, may well indicate that we are starting to uncover the first epoch of galaxy formation.

%%%%% REFERENCES %%%%%%

%\clearpage

% Define journals?
\newcommand{\mnras}{Monthly Notices of the Royal Astronomical Society}
\newcommand{\aap}{Astronomy \& Astrophysics}
\newcommand{\apj}{Astrophysical Journal}
\newcommand{\apjl}{Astrophysical Journal Letters}
\newcommand{\apjs}{Astrophysical Journal Supplement Series}
\newcommand{\araa}{Annual Review of Astronomy and Astrophysics}
\newcommand{\aapr}{Astronomy and Astrophysics Review}
\newcommand{\rmxaa}{Revista Mexicana de Astronomía y Astrofísica}
\newcommand{\pasp}{Publications of the Astronomical Society of the Pacific}

\bibliography{ref}

\begin{thebibliography}{10}

\bibitem{Keres05}
D.~{Kere{\v{s}}}, N.~{Katz}, D.~H. {Weinberg}, R.~{Dav{\'e}}, {\it \mnras\/}
  {\bf 363}, 2 (2005).

\bibitem{Schaye10}
J.~{Schaye}, {\it et~al.\/}, {\it \mnras\/} {\bf 402}, 1536 (2010).

\bibitem{Stark16}
D.~P. {Stark}, {\it \araa\/} {\bf 54}, 761 (2016).

\bibitem{Fernandez16}
X.~{Fern{\'a}ndez}, {\it et~al.\/}, {\it \apjl\/} {\bf 824}, L1 (2016).

\bibitem{Maddox21}
N.~{Maddox}, {\it et~al.\/}, {\it \aap\/} {\bf 646}, A35 (2021).

\bibitem{Wolfe05}
A.~M. {Wolfe}, E.~{Gawiser}, J.~X. {Prochaska}, {\it \araa\/} {\bf 43}, 861
  (2005).

\bibitem{McKinney19}
J.~H. {McKinney}, {\it et~al.\/}, {\it \apj\/} {\bf 874}, 52 (2019).

\bibitem{Shapley03}
A.~E. {Shapley}, C.~C. {Steidel}, M.~{Pettini}, K.~L. {Adelberger}, {\it
  \apj\/} {\bf 588}, 65 (2003).

\bibitem{Finkelstein22}
S.~L. {Finkelstein}, {\it et~al.\/}, {\it \apjl\/} {\bf 940}, L55 (2022).

\bibitem{Finkelstein23}
S.~L. {Finkelstein}, {\it et~al.\/}, {\it \apjl\/} {\bf 946}, L13 (2023).

\bibitem{Bagley23}
M.~B. {Bagley}, {\it et~al.\/}, {\it \apjl\/} {\bf 946}, L12 (2023).

\bibitem{ArrabalHaro23}
P.~{Arrabal Haro}, {\it et~al.\/}, {\it arXiv e-prints\/} p. arXiv:2303.15431
  (2023).

\bibitem{Coe2013}
D.~{Coe}, {\it et~al.\/}, {\it \apj\/} {\bf 762}, 32 (2013).

\bibitem{Hsiao2022}
T.~Y.-Y. {Hsiao}, {\it et~al.\/}, {\it arXiv e-prints\/} p. arXiv:2210.14123
  (2022).

\bibitem{Hsiao23}
T.~Y.-Y. {Hsiao}, {\it et~al.\/}, {\it arXiv e-prints\/} p. arXiv:2305.03042
  (2023).

\bibitem{Tang23}
M.~{Tang}, {\it et~al.\/}, {\it arXiv e-prints\/} p. arXiv:2301.07072 (2023).

\bibitem{Heintz22b}
K.~E. {Heintz}, {\it et~al.\/}, {\it arXiv e-prints\/} p. arXiv:2212.02890
  (2022).

\bibitem{Larson22}
R.~L. {Larson}, {\it et~al.\/}, {\it arXiv e-prints\/} p. arXiv:2211.10035
  (2022).

\bibitem{Fan06}
X.~{Fan}, {\it et~al.\/}, {\it The Astronomical Journal\/} {\bf 132}, 117
  (2006).

\bibitem{MiraldaEscude98}
J.~{Miralda-Escud{\'e}}, {\it \apj\/} {\bf 501}, 15 (1998).

\bibitem{Totani06}
T.~{Totani}, {\it et~al.\/}, {\it Publications of the Astronomical Society of
  Japan\/} {\bf 58}, 485 (2006).

\bibitem{TepperGarcia06}
T.~{Tepper-Garc{\'\i}a}, {\it \mnras\/} {\bf 369}, 2025 (2006).

\bibitem{Buchner14}
J.~{Buchner}, {\it et~al.\/}, {\it \aap\/} {\bf 564}, A125 (2014).

\bibitem{Feroz09}
F.~{Feroz}, M.~P. {Hobson}, M.~{Bridges}, {\it \mnras\/} {\bf 398}, 1601
  (2009).

\bibitem{Leonova22}
E.~{Leonova}, {\it et~al.\/}, {\it \mnras\/} {\bf 515}, 5790 (2022).

\bibitem{Whitler23}
L.~{Whitler}, {\it et~al.\/}, {\it arXiv e-prints\/} p. arXiv:2305.16670
  (2023).

\bibitem{Laursen19}
P.~{Laursen}, J.~{Sommer-Larsen}, B.~{Milvang-Jensen}, J.~P.~U. {Fynbo}, A.~O.
  {Razoumov}, {\it \aap\/} {\bf 627}, A84 (2019).

\bibitem{Hernandez20}
S.~{Hernandez}, {\it et~al.\/}, {\it \apj\/} {\bf 892}, 19 (2020).

\bibitem{Yang17}
H.~{Yang}, {\it et~al.\/}, {\it \apj\/} {\bf 844}, 171 (2017).

\bibitem{Jaskot19}
A.~E. {Jaskot}, T.~{Dowd}, M.~S. {Oey}, C.~{Scarlata}, J.~{McKinney}, {\it
  \apj\/} {\bf 885}, 96 (2019).

\bibitem{Cen15}
R.~{Cen}, T.~{Kimm}, {\it \apjl\/} {\bf 801}, L25 (2015).

\bibitem{Rosdahl22}
J.~{Rosdahl}, {\it et~al.\/}, {\it \mnras\/} {\bf 515}, 2386 (2022).

\bibitem{Yeh23}
J.~Y.~C. {Yeh}, {\it et~al.\/}, {\it \mnras\/} {\bf 520}, 2757 (2023).

\bibitem{Fudamoto22}
Y.~{Fudamoto}, {\it et~al.\/}, {\it \apj\/} {\bf 934}, 144 (2022).

\bibitem{Kennicutt12}
R.~C. {Kennicutt}, N.~J. {Evans}, {\it \araa\/} {\bf 50}, 531 (2012).

\bibitem{Lelli2012}
F.~{Lelli}, M.~{Verheijen}, F.~{Fraternali}, R.~{Sancisi}, {\it \aap\/} {\bf
  537}, A72 (2012).

\bibitem{Hutter21}
A.~{Hutter}, {\it et~al.\/}, {\it \mnras\/} {\bf 503}, 3698 (2021).

\bibitem{Boyett23}
K.~{Boyett}, {\it et~al.\/}, {\it arXiv e-prints\/} p. arXiv:2303.00306 (2023).

\bibitem{Heintz22a}
K.~E. {Heintz}, {\it et~al.\/}, {\it \apjl\/} {\bf 934}, L27 (2022).

\bibitem{Heintz23}
K.~E. {Heintz}, {\it et~al.\/}, {\it \apjl\/} {\bf 944}, L30 (2023).

\bibitem{Naidu20}
R.~P. {Naidu}, {\it et~al.\/}, {\it \apj\/} {\bf 892}, 109 (2020).

\bibitem{Lu23}
T.-Y. {Lu}, {\it et~al.\/}, {\it arXiv e-prints\/} p. arXiv:2304.11192 (2023).

\bibitem{Finkelstein19}
S.~L. {Finkelstein}, {\it et~al.\/}, {\it \apj\/} {\bf 879}, 36 (2019).

\bibitem{McQuinn08}
M.~{McQuinn}, A.~{Lidz}, M.~{Zaldarriaga}, L.~{Hernquist}, S.~{Dutta}, {\it
  \mnras\/} {\bf 388}, 1101 (2008).

\bibitem{Brammer19}
G.~{Brammer}, {Grizli: Grism redshift and line analysis software}, Astrophysics
  Source Code Library, record ascl:1905.001 (2019).

\bibitem{Brammer22}
G.~{Brammer}, {msaexp: NIRSpec analyis tools}, Zenodo (2022).

\bibitem{Planck18}
{Planck Collaboration}, {\it et~al.\/}, {\it \aap\/} {\bf 641}, A6 (2020).

\bibitem{Astropy}
{Astropy Collaboration}, {\it et~al.\/}, {\it \aap\/} {\bf 558}, A33 (2013).

\bibitem{Horne86}
K.~{Horne}, {\it \pasp\/} {\bf 98}, 609 (1986).

\bibitem{Jakobsen22}
P.~{Jakobsen}, {\it et~al.\/}, {\it \aap\/} {\bf 661}, A80 (2022).

\bibitem{astrodrizzle}
S.~{Gonzaga}, W.~{Hack}, A.~{Fruchter}, J.~{Mack}, {\it {The DrizzlePac
  Handbook}\/} (2012).

\bibitem{Carnall18}
A.~C. {Carnall}, R.~J. {McLure}, J.~S. {Dunlop}, R.~{Dav{\'e}}, {\it \mnras\/}
  {\bf 480}, 4379 (2018).

\bibitem{Bruzual03}
G.~{Bruzual}, S.~{Charlot}, {\it \mnras\/} {\bf 344}, 1000 (2003).

\bibitem{Ferland17}
G.~J. {Ferland}, {\it et~al.\/}, {\it \rmxaa\/} {\bf 53}, 385 (2017).

\bibitem{Salim18}
S.~{Salim}, M.~{Boquien}, J.~C. {Lee}, {\it \apj\/} {\bf 859}, 11 (2018).

\bibitem{Kroupa01}
P.~{Kroupa}, {\it \mnras\/} {\bf 322}, 231 (2001).

\bibitem{Iyer19}
K.~G. {Iyer}, {\it et~al.\/}, {\it \apj\/} {\bf 879}, 116 (2019).

\bibitem{Leja19}
J.~{Leja}, A.~C. {Carnall}, B.~D. {Johnson}, C.~{Conroy}, J.~S. {Speagle}, {\it
  \apj\/} {\bf 876}, 3 (2019).

\bibitem{Tacchella23}
S.~{Tacchella}, {\it et~al.\/}, {\it \mnras\/} {\bf 522}, 6236 (2023).

\bibitem{Osterbrock06}
D.~E. {Osterbrock}, G.~J. {Ferland}, {\it {Astrophysics of gaseous nebulae and
  active galactic nuclei}\/} (2006).

\bibitem{Kennicutt98}
J.~{Kennicutt}, Robert~C., {\it \araa\/} {\bf 36}, 189 (1998).

\bibitem{Topping22}
M.~W. {Topping}, {\it et~al.\/}, {\it \mnras\/} {\bf 516}, 975 (2022).

\bibitem{Izotov06}
Y.~I. {Izotov}, G.~{Stasi{\'n}ska}, G.~{Meynet}, N.~G. {Guseva}, T.~X. {Thuan},
  {\it \aap\/} {\bf 448}, 955 (2006).

\bibitem{Sanders23}
R.~L. {Sanders}, A.~E. {Shapley}, M.~W. {Topping}, N.~A. {Reddy}, G.~B.
  {Brammer}, {\it arXiv e-prints\/} p. arXiv:2303.08149 (2023).

\bibitem{Langeroodi22}
D.~{Langeroodi}, {\it et~al.\/}, {\it arXiv e-prints\/} p. arXiv:2212.02491
  (2022).

\bibitem{Nakajima23}
K.~{Nakajima}, {\it et~al.\/}, {\it arXiv e-prints\/} p. arXiv:2301.12825
  (2023).

\bibitem{Curti23}
M.~{Curti}, {\it et~al.\/}, {\it arXiv e-prints\/} p. arXiv:2304.08516 (2023).

\bibitem{Watson11}
D.~{Watson}, {\it \aap\/} {\bf 533}, A16 (2011).

\bibitem{Gordon03}
K.~D. {Gordon}, G.~C. {Clayton}, K.~A. {Misselt}, A.~U. {Landolt}, M.~J.
  {Wolff}, {\it \apj\/} {\bf 594}, 279 (2003).

\bibitem{DeVis19}
P.~{De Vis}, {\it et~al.\/}, {\it \aap\/} {\bf 623}, A5 (2019).

\bibitem{mcmc2013}
D.~{Foreman-Mackey}, D.~W. {Hogg}, D.~{Lang}, J.~{Goodman}, {\it \pasp\/} {\bf
  125}, 306 (2013).

\bibitem{Steinhard22}
C.~L. {Steinhardt}, V.~{Kokorev}, V.~{Rusakov}, E.~{Garcia}, A.~{Sneppen}, {\it
  arXiv e-prints\/} p. arXiv:2208.07879 (2022).

\bibitem{Byler17}
N.~{Byler}, J.~J. {Dalcanton}, C.~{Conroy}, B.~D. {Johnson}, {\it \apj\/} {\bf
  840}, 44 (2017).

\bibitem{Raiter10}
A.~{Raiter}, D.~{Schaerer}, R.~A.~E. {Fosbury}, {\it \aap\/} {\bf 523}, A64
  (2010).

\bibitem{Tanvir19}
N.~R. {Tanvir}, {\it et~al.\/}, {\it \mnras\/} {\bf 483}, 5380 (2019).

\bibitem{Hartoog15}
O.~E. {Hartoog}, {\it et~al.\/}, {\it \aap\/} {\bf 580}, A139 (2015).

\bibitem{Saccardi23}
A.~{Saccardi}, {\it et~al.\/}, {\it \aap\/} {\bf 671}, A84 (2023).

\bibitem{Simcoe20}
R.~A. {Simcoe}, {\it et~al.\/}, {\it arXiv e-prints\/} p. arXiv:2011.10582
  (2020).

\bibitem{Hutter22}
A.~{Hutter}, {\it et~al.\/}, {\it arXiv e-prints\/} p. arXiv:2209.14592 (2022).

\bibitem{Planck20}
{Planck Collaboration}, {\it et~al.\/}, {\it \aap\/} {\bf 641}, A6 (2020).

\bibitem{Nelson19}
D.~{Nelson}, {\it et~al.\/}, {\it \mnras\/} {\bf 490}, 3234 (2019).

\bibitem{Langan20}
I.~{Langan}, D.~{Ceverino}, K.~{Finlator}, {\it \mnras\/} {\bf 494}, 1988
  (2020).

\bibitem{Lovell21}
C.~C. {Lovell}, {\it et~al.\/}, {\it \mnras\/} {\bf 500}, 2127 (2021).

\bibitem{Ucci23}
G.~{Ucci}, {\it et~al.\/}, {\it \mnras\/} {\bf 518}, 3557 (2023).

\bibitem{Curti20}
M.~{Curti}, F.~{Mannucci}, G.~{Cresci}, R.~{Maiolino}, {\it \mnras\/} {\bf
  491}, 944 (2020).

\end{thebibliography}
\bibliographystyle{Science}

% In discussion, also say that Boyett+ notes high Lya and other abs. lines. 

%[Note that Mauerhofer \& Dayal finds $f^{\rm UV}_{\rm esc}\gtrsim 0.9$ at $z\approx 8$ for $M_\star \approx 10^{8.5}$, i.e. large escape expected (we see none).]

% Summary para: Our results indicate XX... These results establish... 

%Other points: 
%- Galaxies younger than $\approx 100$\,Myr will not have had enough time to significantly ionize their surrounding medium. Theoretical and numerical modeling of galaxies at these early times suggest short assembly times of their dark-matter halos and further gas accretion (Mason+15).
%- Two-photon emission is only expected to contribute significantly at very low metallicities (logZ/Zsun $<-2 $)

%%%%% ACKNOWLEDGEMENTS %%%%%%

%\clearpage

\subsection*{Acknowledgements}

We would like to thank Johan P. U. Fynbo and Mark Dickinson for their constructive and enlightening comments on the interpretations of the results presented in this work. We further acknowledge the significant work and effort of the CEERS collaboration in obtaining parts of the observations presented here and are grateful that their early data is publicly available. 
K.E.H. acknowledges support from the Carlsberg Foundation Reintegration Fellowship Grant CF21-0103. A.H. acknowledges support by the VILLUM FONDEN under grant 37459. C.M. acknowledges support by the VILLUM FONDEN under grant 37459 and the Carlsberg Foundation under grant CF22-1322. The Cosmic Dawn Center (DAWN) is funded by the Danish National Research Foundation under grant No. 140. RPN acknowledges funding from JWST programs GO-1933 and GO-2279. Support for this work was provided by NASA through the NASA Hubble Fellowship grant HST-HF2-51515.001-A awarded by the Space Telescope Science Institute, which is operated by the Association of Universities for Research in Astronomy, Incorporated, under NASA contract NAS5-26555.
This work is based on observations made with the NASA/ESA/CSA James Webb Space Telescope. The data were obtained from the Mikulski Archive for Space Telescopes at the Space Telescope Science Institute, which is operated by the Association of Universities for Research in Astronomy, Inc., under NASA contract NAS 5-03127 for JWST. 

\subsection*{Author contributions}

K.E.H. and D.W. drafted the manuscript and led the main analysis. G.B. extracted and reduced the observational data. S.V. performed the IGM and DLA modelling. A.H. provided and interpreted the Astraeus simulations output. V.B.S. performed the SED modelling. All authors contributed to the interpretation and analysis of the results presented in this work. 

\subsection*{Competing interests}

The authors declare no competing interests.

\subsection*{Data and materials availability}

%The data will be made available on Twitter exclusively. 
The JWST imaging and spectroscopic data are publicly available on the JWST MAST archive at \url{https://mast.stsci.edu}. The data has been processed using public software codes; {\tt grizli} v. 1.8.3, \cite{Brammer19} and {\tt MsaExp} v. 0.6.7, \cite{Brammer22}. The reduced data are available from the corresponding author upon reasonable request.

\clearpage

%%%%% SUPPLEMENTARY MATERIALS %%%%%%

\section*{Supplementary Materials}

%Figures to make in the Supps: \\
%1. Laursen+ IGM model for all galaxies. \\
%2. SFR-Mstar and MZ relations, with three LBGs overplotted? \\
%3. Comparison figs with galaxies without DLAs. \\
%4. Emission line fits for the CEERS galaxies. \\
%5. Bagpipes modelling of the CEERS galaxies. \\
%6. NHI distribution vs z. \\
%7. A KS relation plot? \\

\paragraph{Cosmology.}

Throughout this paper we assume the concordance, flat $\Lambda$CDM cosmological model with $H_0 = 67.4$\,km\,s$^{-1}$\,Mpc$^{-1}$, $\Omega_{\rm m} = 0.315$, and $\Omega_{\Lambda} = 0.685$ \cite{Planck18}. Cosmological measurements such as the luminosity distances $d_L$, age of the universe at a given $z$, and transverse proper distances are computed using the cosmology distance calculator from {\tt Astropy} \cite{Astropy}.

\vspace{-0.3cm}

\paragraph{Observations and data reduction.}

This work is mainly based on spectroscopic observations collected through the dedicated JWST Director’s Discretionary time program DD-2750 (PI: P. Arrabal Haro) and the General Observer program GO~1433 (PI: D. Coe). The spectroscopic data obtained of CEERS-43833 and CEERS-16943 (MSA ID's 28 and 1, respectively) have previously been presented in earlier works \cite{ArrabalHaro23}, and likewise for MACS0647-JD \cite{Hsiao23}. For consistency and internal homogeneity, we here reprocess the full spectroscopic dataset, using the custom-made pipeline {\tt MsaExp} v. 0.6.7 \cite{Brammer22}. This code utilizes the Stage 2 output from the MAST JWST archive and performs standard wavelength, flat-field and photometric calibrations on the individual NIRSpec exposure files. This is done with the relevant JWST reference files associated with the CRDS context {\tt jwst1027.pmap}. {\tt MsaExp} further corrects for the $1/f$ noise and the bias levels in individual exposures. We generate the full combined 2D spectra from these output files and optimally extract the 1D spectra using an inverse-weighted sum of the 2D spectra in the dispersion directions \cite{Horne86}. We scale the flux densities of the extracted 1D spectra to match the integrated flux within the available JWST/NIRcam passband in each filter, using a wavelength-dependent polynomial function. This is to improve the absolute flux calibration of the spectra and take into account any potential slit-losses. The required correction factors are typically less than $\approx 10\%$. The NIRSpec prism observations all have wavelength coverage from $0.7\mu$m to $5.3\mu$m, with a varying spectral resolution from $\mathcal{R}\approx 50$ in the blue end to $\mathcal{R}\approx 400$ in the red end \cite{Jakobsen22}. The reduced and photometrically-calibrated spectra of the three galaxies at $z=8.8-11.4$ analyzed here are shown in Fig.~1.  

We further include JWST/NIRCam imaging obtained through the CEERS survey (ERS-1345, PI: S. Finkelstein) \cite{Finkelstein22,Finkelstein23,Bagley23}, in which CEERS-43833 and CEERS-16943 were first identified as a high-redshift galaxy candidates (the latter dubbed Maisie's galaxy \cite{Finkelstein22}). MACS0647-JD was discovered as part of the Cluster Lensing And Supernova survey with Hubble (CLASH) \cite{Coe2013}, and was later observed with JWST/NIRCam (GO~1433, PI: D. Coe). These observations revealed a complex morphology of the multiply-lensed source \cite{Hsiao2022}. For each source, we adopt the photometric measurements from the catalog presented by Brammer et al. (in prep.), currently available online\footnote{\url{https://s3.amazonaws.com/grizli-v2/JwstMosaics/v6/index.html}}. This repository includes reduced images of the raw data obtained from MAST using the public software package {\tt grizli}, which provides astrometric calibrations based on the {\it Gaia}-DR3 catalog, masks imaging artifacts, and drizzles the images to a common pixel scale of $0.\!\!^{\prime\prime}04/$pixel using {\tt astrodrizzle} \cite{astrodrizzle}. The final photometric catalog includes measurements that are based on the most recent JWST/NIRCam photometric zeropoints and have been corrected for extinction due to dust in the Galaxy. For the modelling of the galaxy SEDs we use the aperture-matched photometry from this catalog, derived using a circular aperture with diameter $0.\!\!^{\prime\prime}5$.

\vspace{-0.3cm}

\paragraph{Morphology and galaxy sizes.}

We adopt the size measurements of MACS0647-JD from \cite{Hsiao23}, assuming that the galaxy spans the edges of the A and B components, yielding an effective diameter of $\sim 0.5$\,kpc. %The imaging data of CEERS-43833 and CEERS-16943 have also previously been presented \cite{Finkelstein22,ArrabalHaro23}, but for consistency we here measure the spatial extent of these two galaxies at rest-frame UV wavelengths. 
For CEERS-43833 we measure full-width-half-maximum (FWHM) angular major and minor axes of $0.\!\!^{\prime\prime}21 \times 0.\!\!^{\prime\prime}19$ (in F200W, i.e. rest-frame UV), corresponding to $0.97\times 0.88$\,kpc at $z=8.7622$. For CEERS-16943, we measure FWHM angular major and minor axes of $0.\!\!^{\prime\prime}17 \times 0.\!\!^{\prime\prime}14$ (F200W), corresponding to $0.66\times 0.54$\,kpc at $z=11.409$. We estimate the physical sizes enclosing half of the total rest-frame UV luminosity estimated from the Gaussian modelling (half-light semi axes = $a_{\rm maj,min}$ = FWHM$_{\rm maj,min}/2$), yielding effective areas of the projected ellipses of $0.67$\,kpc$^{-2}$ (CEERS-43833) and $0.28$\,kpc$^{-2}$ (CEERS-16943), respectively.

\vspace{-0.3cm}

\paragraph{SED modelling.}

We adopt the outputs from the SED modelling of the physical properties of the MACS0647-JD complex as a whole, as presented in \cite{Hsiao23}, and summarized in Table~\ref{tab:galprops}. Their analysis reveals a low-mass, $\log (M_\star / M_\odot) = 8.1\pm 0.3$, galaxy with SFR$_{\rm SED} = 8\pm 3\,M_\odot\,{\rm yr}^{-1}$, and a blue continuum slope with $\log U = -1.9\pm 0.2$ and $A_{\rm V,SED} = 0.07\pm 0.06$\,mag. 
To infer the physical properties of CEERS-43833 and CEERS-16943, we model their SEDs using the Python code Bayesian Analysis of Galaxies for Physical Inference and Parameter EStimation ({\tt Bagpipes}) \cite{Carnall18}, which incorporates the stellar population models from Bruzual \& Charlot \cite{Bruzual03}. We perform spectrophotometric fitting of the two galaxy SEDs, fixing the redshifts to $z_{\rm spec}$ inferred from the nebular emission lines. The emission lines are added to the model spectra in {\tt Bagpipes} using a grid computed from {\tt Cloudy} models \cite{Ferland17}. We allow the ionization parameter, $U$, to vary between $-3 < \log U < -1$ and assume the attenuation curve from Salim et al. \cite{Salim18} to account for the typically higher ionization parameters and the more steep attenuation curves of high-$z$ galaxies. Throughout, we assume a Kroupa initial mass function (IMF) \cite{Kroupa01}. The best-fit SED models for CEERS-43833 and CEERS-16943 are shown in Fig.~S1. 

For simple, constant SFHs we infer young ($\lesssim 100$\,Myr) stellar populations on average in both galaxies. We caution, however, that the inferred ages are heavily dependent on the assumed SFH parameterization. 
The accurate spectrophotometric fitting allow us to incorporate more flexible, non-parametric SFHs for each galaxy which more accurately capture the stellar ages and mass build-up \cite{Iyer19}. This effectively increases the inferred stellar masses (by 0.2--0.3\,dex for these two galaxies) compared to the best-fit models assuming constant SFHs, as also demonstrated in previous works \cite{Leja19,Tacchella23}. This parameterization fits for quartiles of total mass formed, and includes a prolonged, less active period of star formation to take into account possible hidden mass from an older stellar population due to ``outshining'' from a potential younger, burstier star-forming population. Table~1 reports the median and 16th and 84th percentiles from the resulting posterior distributions on the stellar mass, SFR, dust attenuation, and ionization parameter for each galaxy. The results for CEERS-16943 are overall consistent with those obtained by Finkelstein et al. \cite{Finkelstein22}, which were, however, based on the photometric measurements only. 

\vspace{-0.3cm}

\paragraph{Emission line measurements.}

The emission line detections and spectroscopic redshift for CEERS-43833 and MACS0647-JD have previously been reported based on several nebular emission lines \cite{ArrabalHaro23,Hsiao23}, but for consistency we here rederive the line fluxes and redshifts. We measure $z_{\rm spec}=8.7622\pm 0.0002$ and $z_{\rm spec} = 10.170\pm 0.003$, respectively, as listed in Table~1, see also Fig.~S2. The line fluxes are derived by jointly modelling the local continuum and the emission lines represented by Gaussian profiles, tying the redshift and width of the lines across transitions, but allow the individual line fluxes to vary. The measured line fluxes are summarized in Table~S1. 

The spectroscopic redshifts of CEERS-16943 have also previously been reported \cite{ArrabalHaro23}, however, only based on the approx. onset of the \lya\ break. As demonstrated in this work, this will overestimate the redshift due to broadening of \lya\ from the strong DLA feature. We here determine the spectroscopic redshift independently and provide the line fluxes for the new detected emission lines in the spectrum. We measure $z=11.409\pm 0.001$ for CEERS-16943 based on the detection of [\oii]\,$\lambda 3727$, [Ne\,{\sc iii}]\,$\lambda 3870$, and He\,{\sc i}\,$\lambda 3889$, see Fig.~S1. This is the to-date most distant detection of nebular emission lines and thus also the earliest detection of single metals or elements heavier than hydrogen. We note that in the previous analysis of CEERS-16943 \cite{ArrabalHaro23} the feature at the location of [\oii] was proposed to be caused by image defects. However, based on our independent reduction and the detection of two other features at the same redshift, we argue that this emission line feature is real. 

\vspace{-0.3cm}

\paragraph{Physical Properties.} 

In the typical case B recombination scenario, we expect Balmer line ratios ${\rm H}\gamma/{\rm H}\beta = 0.47$ and ${\rm H}\delta/{\rm H}\beta = 0.26$ for electron temperatures $T_e = 10^4$\,K \cite{Osterbrock06}, which is consistent within $1\sigma$ of our measurements ${\rm H}\gamma/{\rm H}\beta  = 0.48\pm 0.08$ and ${\rm H}\delta/{\rm H}\beta = 0.21\pm 0.07$ for CEERS-43833. This implies negligible extinction due to dust, as also supported by the SED modelling. For MACS0647-JD and CEERS-16943 we are not able to measure the Balmer decrement, but their SEDs imply similar low dust attenuation. 

For CEERS-43833, we derive the SFR from H$\beta$ \cite{Kennicutt98}, where SFR = $5.5\times 10^{-42}L_{\rm H\beta} ({\rm erg/s}) \times f_{\rm H\alpha/H\beta}$, assuming a Kroupa IMF \cite{Kroupa01}. Here, $f_{\rm H\alpha/H\beta} = 2.87$ is the predicted ratio from the Case B recombination scenario at $T_e = 10^{4}\,$K. For MACS0647-JD and CEERS-16943, where H$\beta$ is redshifted outside of the JWST/NIRSpec wavelength coverage, we use the relation between SFR and [\oii]\,$\lambda 3727$ line luminosity \cite{Kennicutt98}; SFR = $1.0\times 10^{-41}L_{\rm [O\,\textsc{ii}]} ({\rm erg/s})$, again corrected to a Kroupa IMF. 
The exact choice of the IMF introduces a 0.3\,dex systematic uncertainty on these measurement, which we include in our derivation. The SFRs range from 1 to 15\,$M_\odot$\,yr$^{-1}$, and are summarized in Table~1.
Combined with the stellar mass inferred from the SED fitting, we compute specific SFRs in the range log(sSFR/yr$^{-1}) = -7.55\pm 0.30$ (CEERS-43833) to log(sSFR/yr$^{-1}) = -8.25\pm 0.30$ (CEERS-16943). These are consistent with the typical star-forming galaxy population in the reionization epoch \cite{Topping22,Heintz22b}. 

We infer the gas-phase metallicities for each galaxy based on strong-line diagnostics, since we do not detect the auroral [\oiii]\,$\lambda4363$ emission line enabling direct $T_e$-based measurements \cite{Izotov06}, except for in the spectrum of MACS0647-JD (see also \cite{Hsiao23}). This spectrum, however, does not cover the [\oiii]\,$\lambda\lambda 4960,5008$ transition doublet as well, required to determine the metallicity through the direct $T_e$-method. We use the recent set of strong-line calibrations based on the direct metallicities inferred from a large sample of galaxies at $z=2-9$ spectroscopically observed with JWST \cite{Sanders23}. For CEERS-43833, we report the metallicity inferred using the O3 = [\oiii]\,$\lambda 5008$/H$\beta$ line ratio, $12+\log$(O/H) $= 7.46\pm 0.09$, as this calibration is less affected by dust extinction and the ionization state of the gas, in addition to any potential issues with the flux calibration as a function of wavelength. Further, it shows the least intrinsic scatter as derived by Sanders et al. \cite{Sanders23}. For MACS0647-JD and CEERS-16943, where these features are redshifted out of the NIRSpec wavelength coverage, we instead adopt the Ne3O2 = [Ne\,\textsc{iii}]\,$\lambda 3870$ /  [\oii]\,$\lambda 3727$ line ratio, yielding $12+\log$(O/H) $= 7.40\pm 0.3$ and $7.70\pm 0.30$, respectively. The inferred metallicity of MACS0647-JD is consistent with previous estimates \cite{Hsiao23}, based on various methods and line diagnostics. Using the Ne3O2 calibration to infer the metallicity of CEERS-43833 yields a consistent result to the O3 calibration within the uncertainties. The inferred metallicities are in the range $12+\log$(O/H) = 7.40 - 7.70, i.e. 5-10\% solar, as listed in Table~1. In Fig.~S3, we show the physical properties of the three young, $z=9-11$ galaxies, compared to the typical population of star-forming galaxies at $z=7-10$ \cite{Heintz22b}, in terms of their SFRs, stellar masses, and gas-phase metallicities. Their inferred physical properties are consistent with the established mass-metallicity (MZ) relation at $z>7$ \cite{Langeroodi22,Heintz22b,Nakajima23}, and furthers shows a significant offset by 0.3--0.5\,dex at $2-5\sigma$ from the fundamental-metallicity relation (FMR) at $z\approx 0$ as also observed for other star-forming galaxies at $z>7$ \cite{Heintz22b,Nakajima23,Curti23}. 

A limit on the metallicity of the DLAs can be determined from their lack of significant dust extinction. At a Milky Way dust-to-gas (DTG) ratio \cite{Watson11}, such DLAs would produce approximately 50 magnitudes of extinction at the \lya\ wavelength using an SMC-like extinction curve \cite{Gordon03}. Using the inferred galaxy attenuation as an estimate of the extinction, we find DTG ratios consistent with the power-law relation of \cite{DeVis19} for our inferred emission-line metallicities. This also allows us partially to address the question of whether such extreme DLAs do exist at lower redshift but are not detected due to dust extinction. Assuming such galaxies could be detected with at most one magnitude of extinction at \lya\ and a column density of $N_{\rm H\,\textsc{i}}\simeq 10^{22}$\,cm$^{-2}$, this corresponds to DTG ratios $\lesssim 2.5\times 10^{-4}$. Such low DTG ratios are only observed in galaxies with metallicities of 12+log(O/H) $ \lesssim 8.15$. This suggests that the absence of these extreme DLAs at lower redshift is real and due to a lack of huge foreground \hi\ reservoirs rather than dust bias.

\vspace{-0.3cm}

\paragraph{The damped \lya\ absorption feature.}

We model the broad DLA features in each galaxy at $z=9-11$ by jointly fitting the optical depth from the Gunn-Peterson trough due to absorption from \hi\ in the IGM $\tau_{\rm IGM}(z)$ and the \hi\ located in the immediate surroundings of the galaxies, $\tau_{\rm ISM}$, as described in the Main Text. This allows us to independently model the neutral hydrogen fraction, $x_{\rm H\,\textsc{i}}$, in the IGM and the \hi\ column density in the near-proximity to the galaxies, most likely in the interstellar or circumgalactic medium. We use a set of galaxy template spectra designed to match the blue rest-frame UV colors of galaxies at $z>8$ \cite{Larson22} as the intrinsic continuum galaxy models. We leave the visual extinction $A_V$, the average IGM neutral fraction, $x_{\rm H\,\textsc{i}}$, and the \hi\ column density, $N_{\rm H\,\textsc{i}}$, at $z=z_{\rm gal}$ as free parameters. We recover extremely strong DLAs in all three $z=9-11$ galaxies, with column densities $N_{\rm H\,\textsc{i}} > 10^{22}\,{\rm cm^{-2}}$. The modelling is largely insensitive to $x_{\rm H\,\textsc{i}}$, as the DLA feature dominates the observed line profile in the three sources (see Fig.~1). From the modelling, we infer lower bounds on $x_{\rm H\,\textsc{i}}$ from the 68\% and 95\% highest density intervals of the posterior distributions, $x_{\rm H\,\textsc{i}} = 0.35-0.4$ and $x_{\rm H\,\textsc{i}} = 0.06-0.08$, respectively. 

To verify the robustness of the models, we also run a Markov Chain Monte Carlo (MCMC) algorithm \cite{mcmc2013} on the spectra. This fitting method finds similar values for the neutral hydrogen fraction, $x_{\rm H\,\textsc{i}}$, of the IGM and for the \hi\ column density. We further consider an expanded set of intrinsic galaxy models, including models with bursty SFHs and continous SFHs on longer timescales (10--30\,Myr) from \cite{Larson22}, and models with more physically motivated IMFs for galaxies at $z=8-12$ \cite{Steinhard22}. The variations in these intrinsic galaxy models introduce at most a systematic uncertainty of 0.1\,dex on the derived \hi\ column densities, which we propagate with the statistical uncertainty to the total uncertainty listed in Table~1. This is because the strong DLA feature dominates the spectral shape around \lya\ over the more subtle variations in the various galaxy continuum emission models. 

To exclude any potential false-positive detections of DLAs due to instrumental issues with the flux or wavelength calibration of NIRSpec, we compare CEERS-43833 to a set of other high-$z$ galaxies in Fig.~S2 with similar spectroscopic observations (from \cite{Heintz22b}). These comparison galaxies are selected to be at similar redshifts, with stellar masses and metallicities ranging from $M_\star = 10^{8.6} - 10^{10}\,M_\odot$ and $12+\log({\rm O/H}) = 7.4 - 8.0$, and have equally high S/N in the spectral regions around \lya. We convert and scale the spectroscopic redshifts to $z=8.7622$ and normalize the continuum for each galaxy to that of CEERS-43833 to improve the visual comparison. It is evident that none of these comparison galaxies exhibit similar broadening of the \lya\ trough, beyond that expected for the IGM at $x_{\rm H\,\textsc{i}} = 0.1-1$. In Fig.~S4 we also overplot the various IGM models convolved with the wavelength-dependent spectral resolution of the JWST/NIRSpec prism observations. While the IGM models are less easily discerned, the broad DLA feature is still readily recovered even in the low spectral-resolution observations. This excludes the broad \lya\ feature as being induced by instrumental broadening effects.  

We also consider other physical effects that might mimic the broad \lya\ troughs observed in the galaxy spectra. For instance, nebular continuum emission models with varying stellar mass ranges have been shown to produce similar damped \lya\ profiles \cite{Byler17} as observed. However, this effect is only recovered for stellar populations dominated by older, low-mass ($<5\,M_\odot$) stars in these models, which we argue are unlikely to be the case for the three high-$z$ galaxies studied here, and is also inconsistent with the inferred blue spectral slopes and young stellar populations. Further, 2$\gamma$ continuum emission may potentially imprint a distinct continuum feature near \lya\ as well \cite{Raiter10}. However, the probability of these transitions only becomes significant in extremely metal-poor galaxies (less than $\times 10$ the metallicities of the $z=9-11$ galaxies), where the physical state of the star-forming gas would depart from Case B, so we argue that this mechanism is also unlikely to explain the observed broad \lya\ troughs.

The extended DLAs covering the UV emission of the three young, $z=9-11$ galaxies are the strongest observed to-date in emission-selected star-forming galaxies, considering some of the most metal-poor galaxies at $z\approx 0$ \cite{McKinney19,Hernandez20} or Lyman-break galaxies at $z\approx 3$ \cite{Shapley03}. Further, only about $\lesssim 15\%$ of $\gamma$-ray burst absorption systems at redshifts $z=2-6$ show \hi\ column densities exceeding $N_{\rm H\,\textsc{i}} > 10^{22}\,{\rm cm^{-2}}$ \cite{Tanvir19}. Moreover, the most distant DLAs previously detected have been observed out to redshifts $z\approx 6$ in $\gamma$-ray burst absorbers \cite{Hartoog15,Saccardi23}, and indirectly at $z=6.84$ in a single quasar absorption system \cite{Simcoe20}. For Lyman-break galaxies, the DLA feature has only been seen out to $z\approx 3$ \cite{Shapley03} due to the lacking NIR sensitivity of ground-based telescopes. With these JWST observations, we are now able to push this observational barrier and detect DLAs out to redshifts beyond $z>11$, i.e. within 400\,Myr after the Big Bang at almost a factor of two larger redshifts than previously possible. In Fig.~S5 we compare the \hi\ column densities of the $z=9-11$ galaxies to representative samples of metal-poor, $z\approx 0$ ``high-$z$ analogs'' \cite{McKinney19,Hernandez20} and the most recent collection of $\gamma$-ray burst absorbers expanded upon the initial sample presented in \cite{Tanvir19}. We further mark the maximum \hi\ column densities based on the lowest \lya\ equivalent width (EW) sources in the $z\approx 3$ Lyman-break galaxy sample by \cite{Shapley03}, assuming that $N_{\rm H\,\textsc{i}} = 1.88\times 10^{18} \, {\rm EW}^2_{\rm Ly\alpha}$\,cm$^{-2}$ in the \lya\ damping regime. 

\vspace{-0.3cm}

\paragraph{IGM modelling.}

To include a more physically motivated transmission of the IGM, we consider a typical sightline from a galaxy at $z \simeq 8.8$ (matching approximate CEERS-43833), based on high-resolution cosmological hydro-simulations, post-processed with radiative transfer of ionizing UV and Ly$\alpha$ photons \cite{Laursen19}.
The transmission $T(\lambda)$ as a function of wavelength is calculated as the median value (and 16/84 and 5/95 percentiles) of $10^4$ sightlines emanating from $\sim 300$ galaxies in a zoomed-in region, 27 comoving Mpc across, resolved down to 10 pc scales.
Because of the large neutral fraction of the IGM, the scattering of Ly$\alpha$ photons \emph{into} the line of sight cannot be neglected out to several times the virial radius of a galaxy, and hence the sightlines start at $10 r_\mathrm{vir}$. At this point the radiative transfer is dominated by scattering \emph{out of} the line of sight.
More details on both the hydro-simulations and the radiative transfer can be found in \cite{Laursen19}.

In Fig.~S6, we show the median and the ``1$\sigma$'' (given by the 16th and the 84th percentiles) and the ``2$\sigma$'' (given by the 5th and the 95th percentiles) distributions of these transmission curves shifted to the redshift of the galaxy at $z = 8.7622$.
Evidently, none of these modelled galaxy sightlines are able to reproduce the large Ly$\alpha$ damping observed in this particular galaxy spectrum, supporting a large abundance of \hi\ in the interstellar or circumgalactic medium of the galaxy. We caution, however, that the models in \cite{Laursen19} assume a relatively low IGM neutral gas fraction, $x_{\rm H\,\textsc{i}} = 0.13$, so the \lya\ transmission curves should effectively be treated as lower limits for these particular models.

\vspace{-0.3cm}

\paragraph{Astraeus simulation.}

The {\tt Astraeus} simulations included in this work, couples a semi-analytical galaxy evolution model to a semi-numerical reionization scheme and runs on the outputs of a dark-matter-only $N$-body simulation. It follows the key baryonic processes of gas accretion, gas and stellar mass being brought in by mergers, star formation, supernovae feedback, metal and dust enrichment, and radiative feedback from reionization, as well as the spatial ionization of the IGM (see \cite{Hutter21, Hutter22} for details). In this paper, we analyse the {\sc mhdec} simulation (described in \cite{Hutter22}), which assumes the escape fraction of ionising photons to decrease with rising halo mass. This simulation reproduces the observed UV luminosity and stellar mass functions at $z=5-10$ as well as the electron optical depth and the IGM's neutral hydrogen fraction at $z<6$ inferred from Planck \cite{Planck20} and quasar absorption line measurements, respectively.

%Overall, this galaxy is thus a ``normal'' star-forming Lyman-break galaxy in the reionization at $z\approx 9$. 

\clearpage

%%%%% SUPPLEMENTARY FIGURES %%%%%%

\noindent {\bf Table S1. Line flux measurements reported in units of $10^{-19}$\,erg\,s$^{-1}$\,cm$^{-2}$\,\AA$^{-1}$.} The measurements for MACS0647-JD have not been corrected for the magnification factor $\mu = 8\pm 1$. 

\begin{table}[!h]
    \centering
%    \caption{{\bf Line flux measurements reported in units of $10^{-19}$\,erg\,s$^{-1}$\,cm$^{-2}$\,$\AA^{-1}$.}}
    \begin{tabular}{lccc}
    \hline
    Transition & CEERS-43833 & MACS0647-JD & CEERS-16943\\
    \hline
%    \vspace{0.2cm}
 %       \siv\ & -- & XX \\ 
 %       \civ\ & -- & XX \\
        $[{\rm O\,\textsc{ii}}]\,\lambda 3727$ & $5.1\pm 0.6$ & $2.3\pm 0.8$ & $2.4\pm 0.6$ \\
        $[{\rm Ne\,\textsc{iii}}]\,\lambda 3870$ & $4.5\pm 0.6$ & $2.9\pm 0.8$ & $1.9\pm 0.5$ \\
        \hei\,$\lambda 3889$ & -- & $1.7\pm 0.8$ & $1.6\pm 0.5$ \\
        H$\delta$ & $2.3\pm 0.6$ & $2.0\pm 0.8$ & -- \\
        H$\gamma$ & $4.6\pm 0.5$ & $4.4\pm 0.8$ & -- \\
        H$\beta$ & $9.2\pm 0.5$ & -- & -- \\
        $[{\rm O\,\textsc{iii}}]\,\lambda 4363$ & -- & $2.1\pm 0.7$  & -- \\
        $[{\rm O\,\textsc{iii}}]\,\lambda 4960$ & $17.1\pm 0.6$ & -- & -- \\
        $[{\rm O\,\textsc{iii}}]\,\lambda 5008$ & $51.4\pm 0.6$ & -- & -- \\
        \hline
    \end{tabular}
    \label{tab:lines}
\end{table}

%\noindent {\bf Table S1.} Line flux measurements and equivalent widths, reported in units of $10^{-19}$\,erg\,s$^{-1}$\,cm$^{-2}$\,$\AA^{-1}$ and $\AA$, respectively.

\clearpage

\begin{figure}[!ht]
    \centering
    \includegraphics[width=0.7\textwidth]{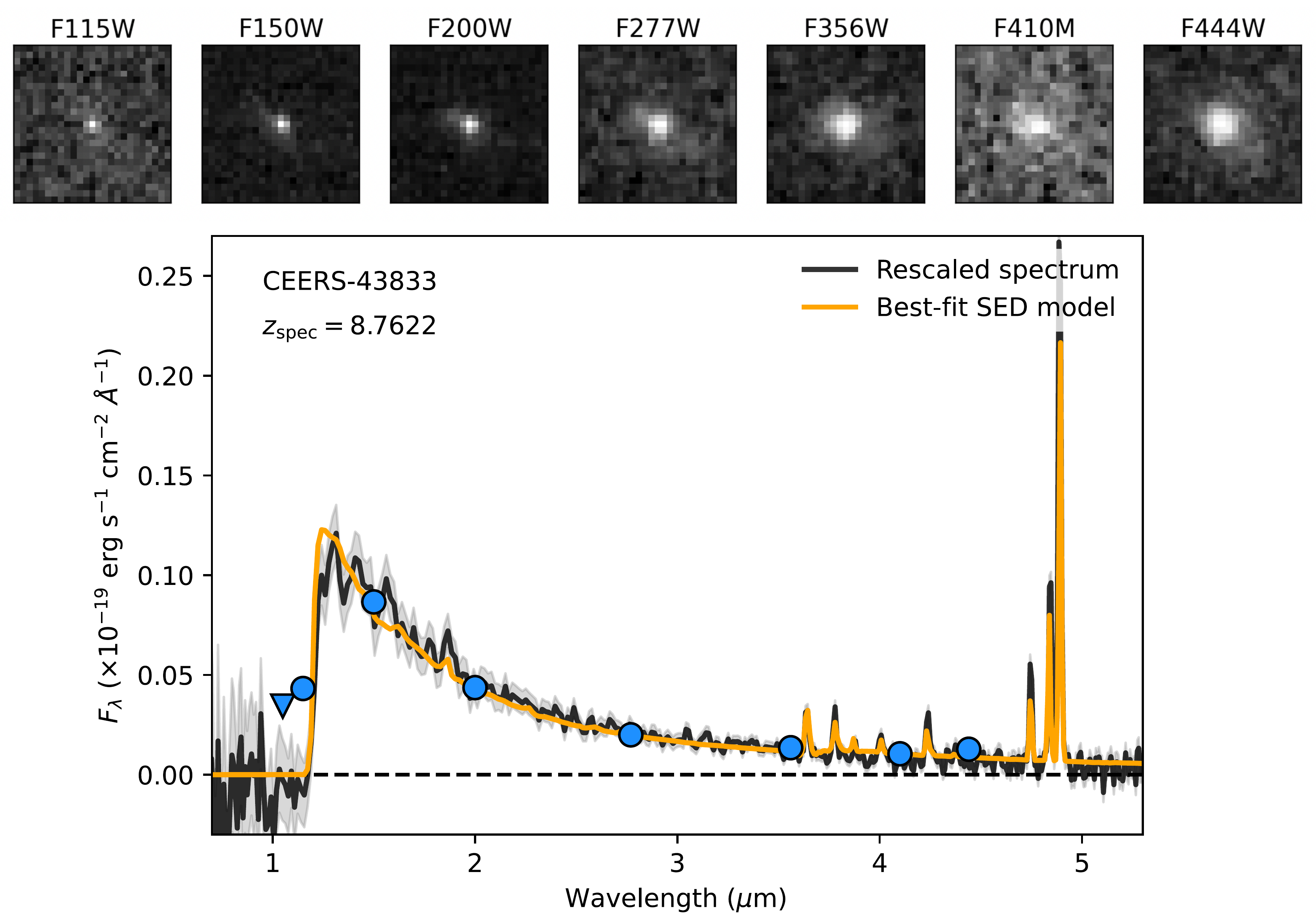}
    \includegraphics[width=0.7\textwidth]{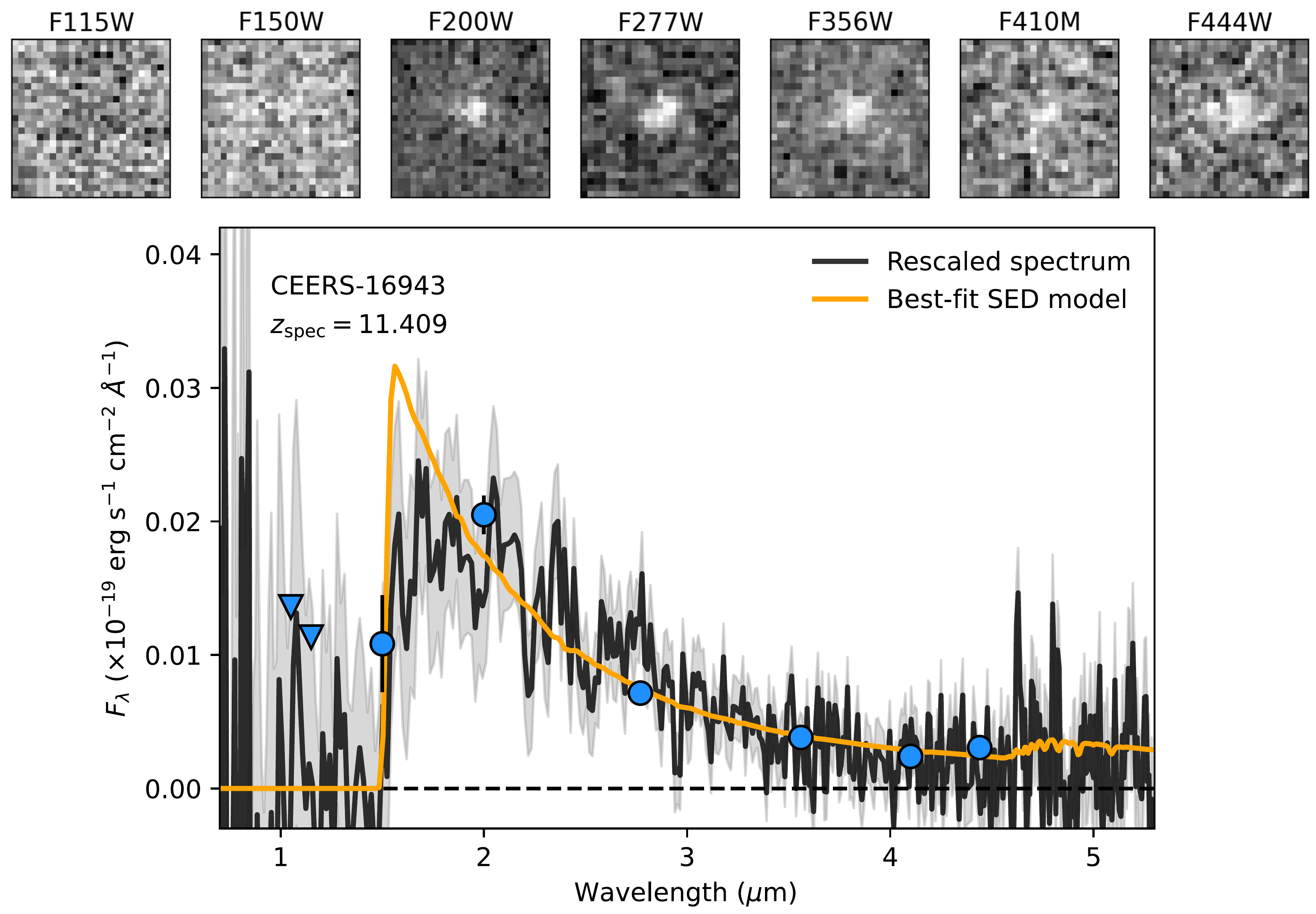}
%    \caption{}
    \label{fig:figs1}
\end{figure}

\noindent {\bf Fig. S1. Imaging and best-fit SED models.} In the top panels for CEERS-43833 and CEERS-16943 are shown the reduced, drizzled images of the source in each available JWST/NIRCam filter. In the bottom panels are shown the reduced and photometrically-calibrated 1D spectra (black curves) with the associated error spectra (grey-shaded regions). The best-fit SED models from the spectrophotometric fitting using {\tt Bagpipes} are overplotted in orange. 

\clearpage

\begin{figure}[!ht]
    \centering
    \includegraphics[width=0.7\textwidth]{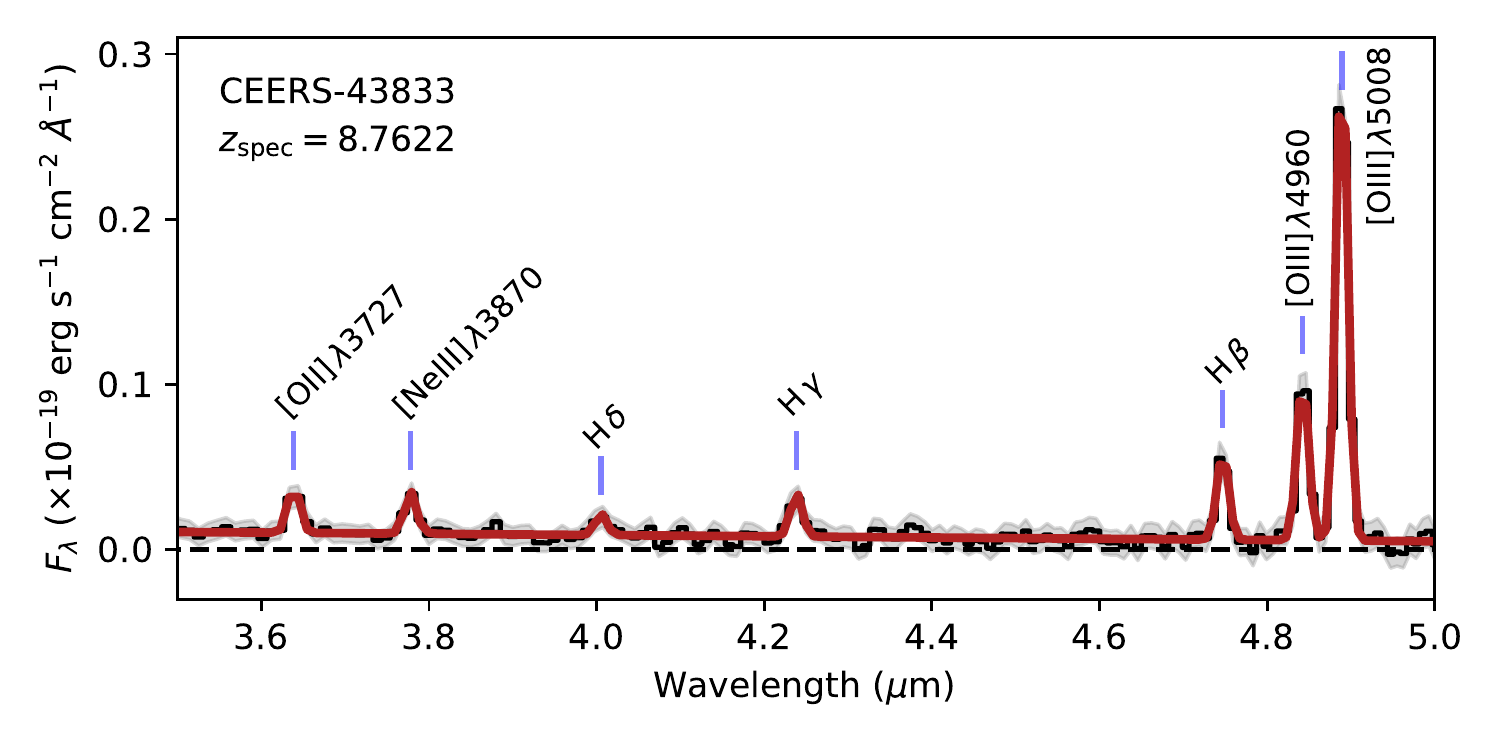}
    \includegraphics[width=0.7\textwidth]{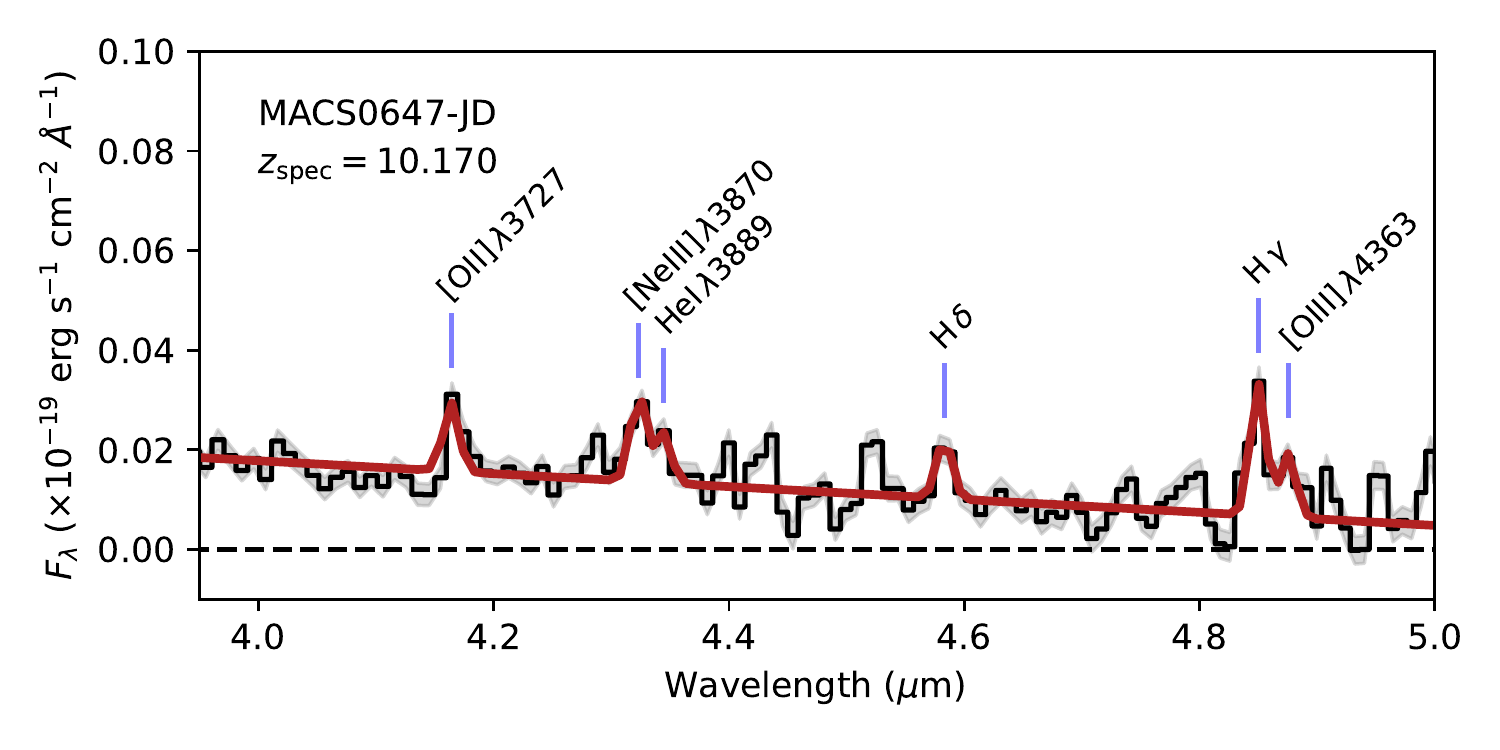}
    \includegraphics[width=0.7\textwidth]{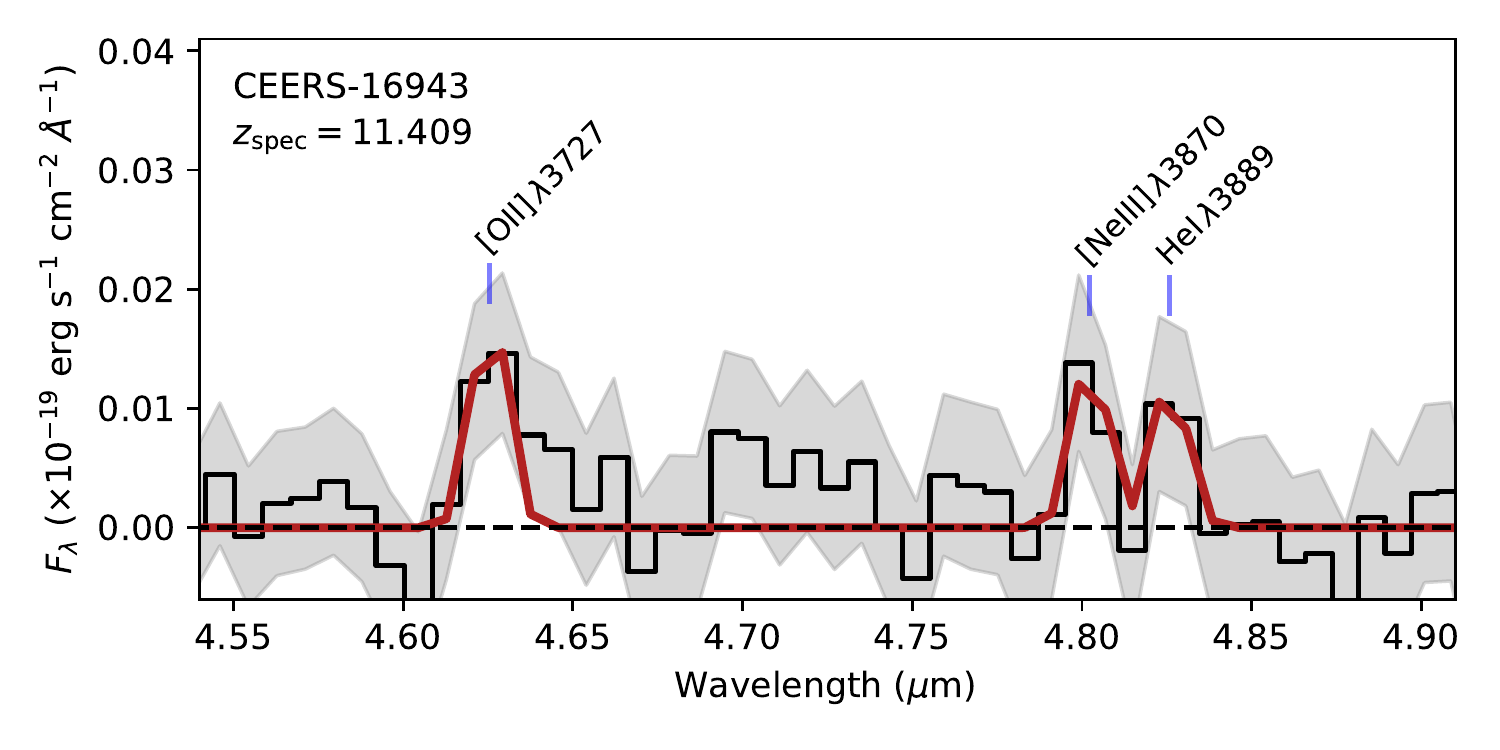}
%    \caption{}
    \label{fig:figs2}
\end{figure}

\noindent {\bf Fig. S2. Detected line emission features.} The reduced and photometrically-calibrated 1D spectra for each galaxy at $z=9-11$ (marked in the top left of each panel) are shown in black with the associated error spectra shown by the grey-shaded regions. The best-fit local continuum emission and Gaussian line profiles are overplotted in red. 

\clearpage

\begin{figure}
    \centering
    \includegraphics[width=\textwidth]{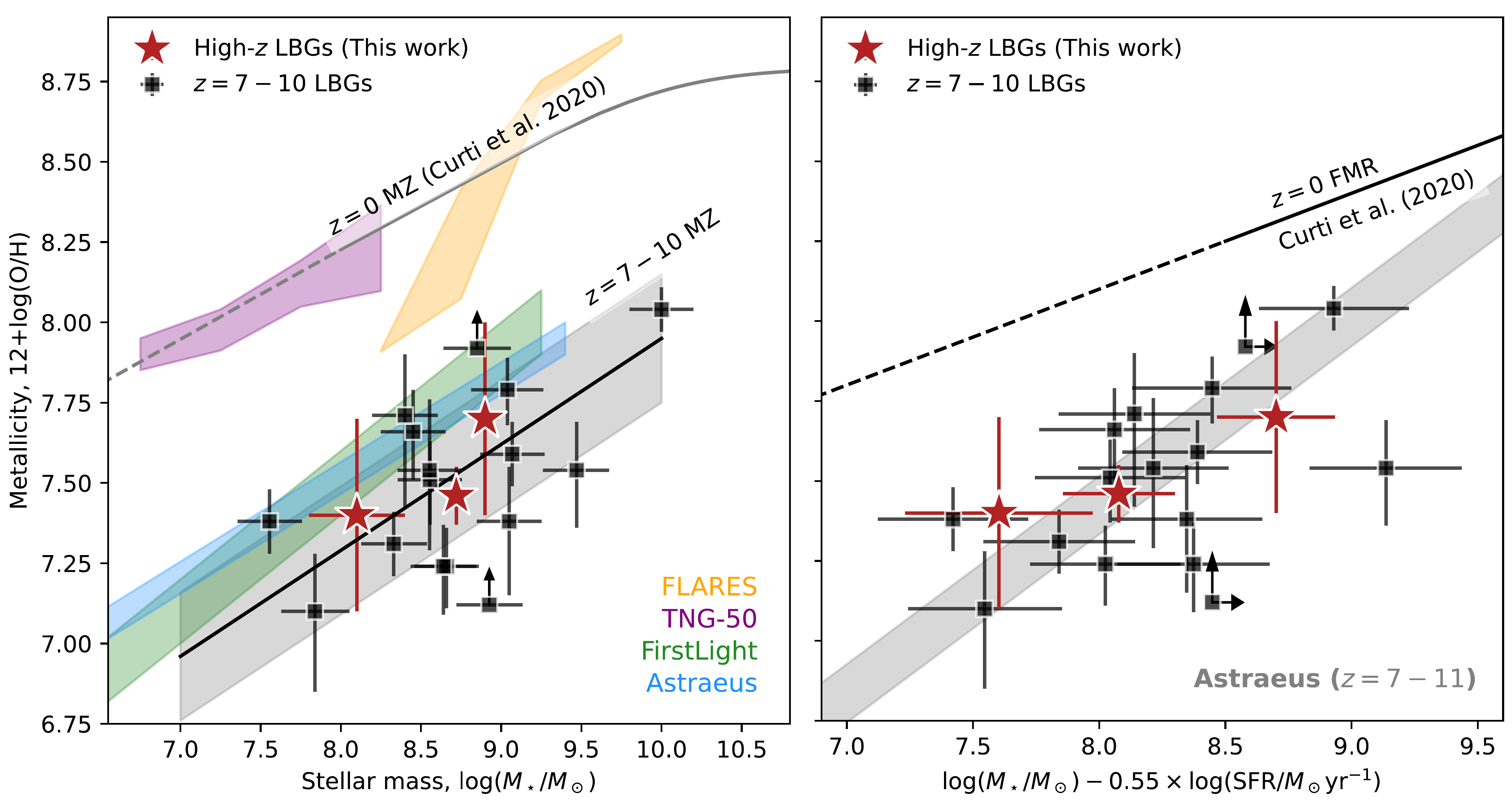}
%    \caption{}
    \label{fig:figs3}
\end{figure}

\noindent {\bf Fig. S3. Physical properties of the young, $z=9-11$ star-forming galaxies.} In the left panel, the inferred stellar masses and gas-phase metallicities derived in this work (red star symbols) are compared to individual observations (black) and the best-fit mass-metallicity relation of main-sequence star-forming galaxies at $z=7-10$ \cite{Heintz22b}, in addition to a suite of predictions from simulations at similar redshifts \cite{Nelson19,Langan20,Lovell21,Ucci23}, as indicated by the labels in the bottom right. In the right panel, is shown the fundamental-metallicity relation again including individual star-forming galaxies at $z=7-10$ \cite{Heintz22b}, and predictions from the {\tt Astraeus} simulation at similar redshifts \cite{Ucci23}. Both panels show the best-fit mass-metallicity (left) and fundamental-metallicity relation (right) of galaxies at $z\approx 0$ \cite{Curti20} for comparison.

\clearpage

\begin{figure}
    \centering
    \includegraphics[width=0.49\textwidth]{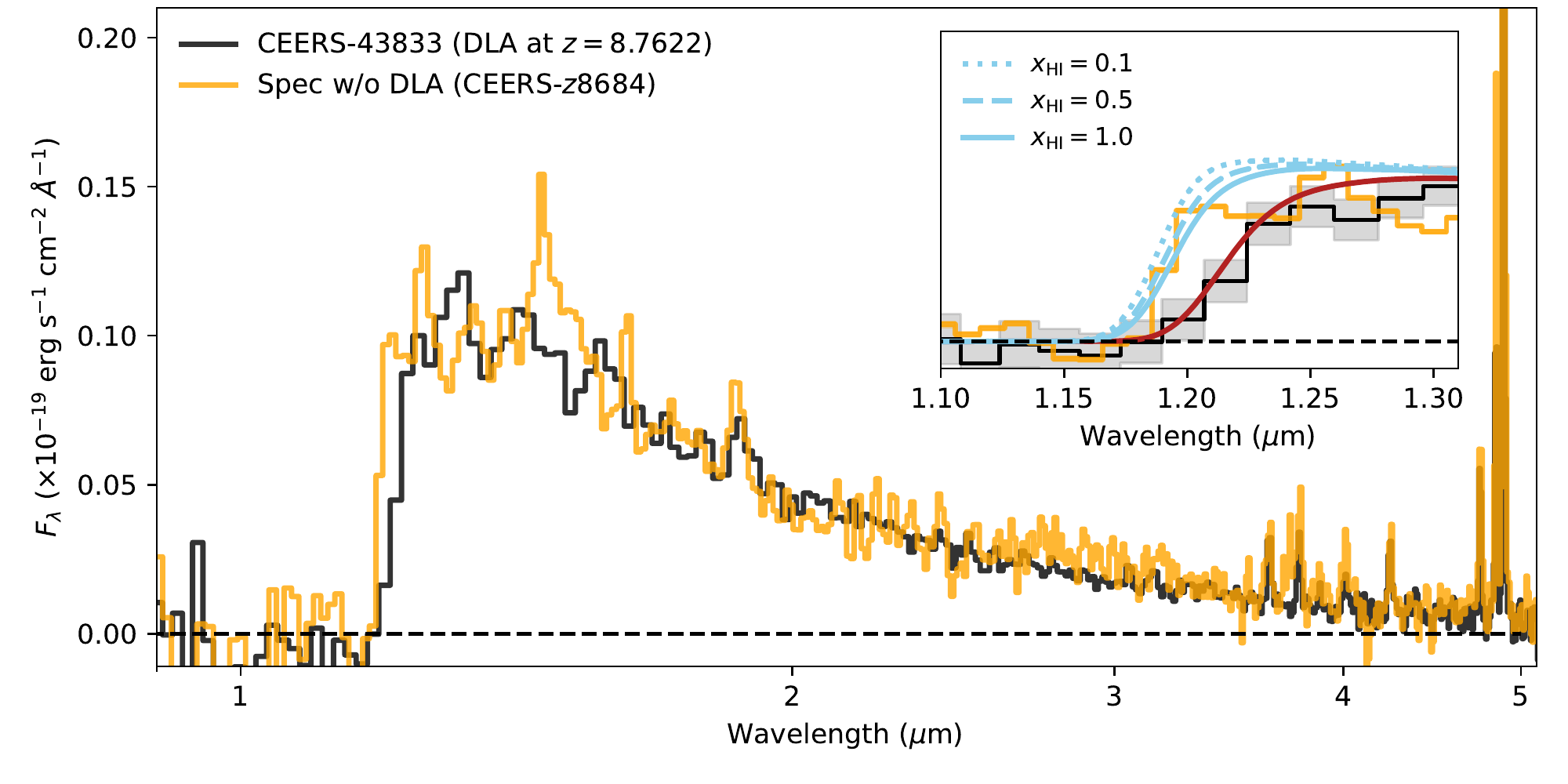}
    \includegraphics[width=0.49\textwidth]{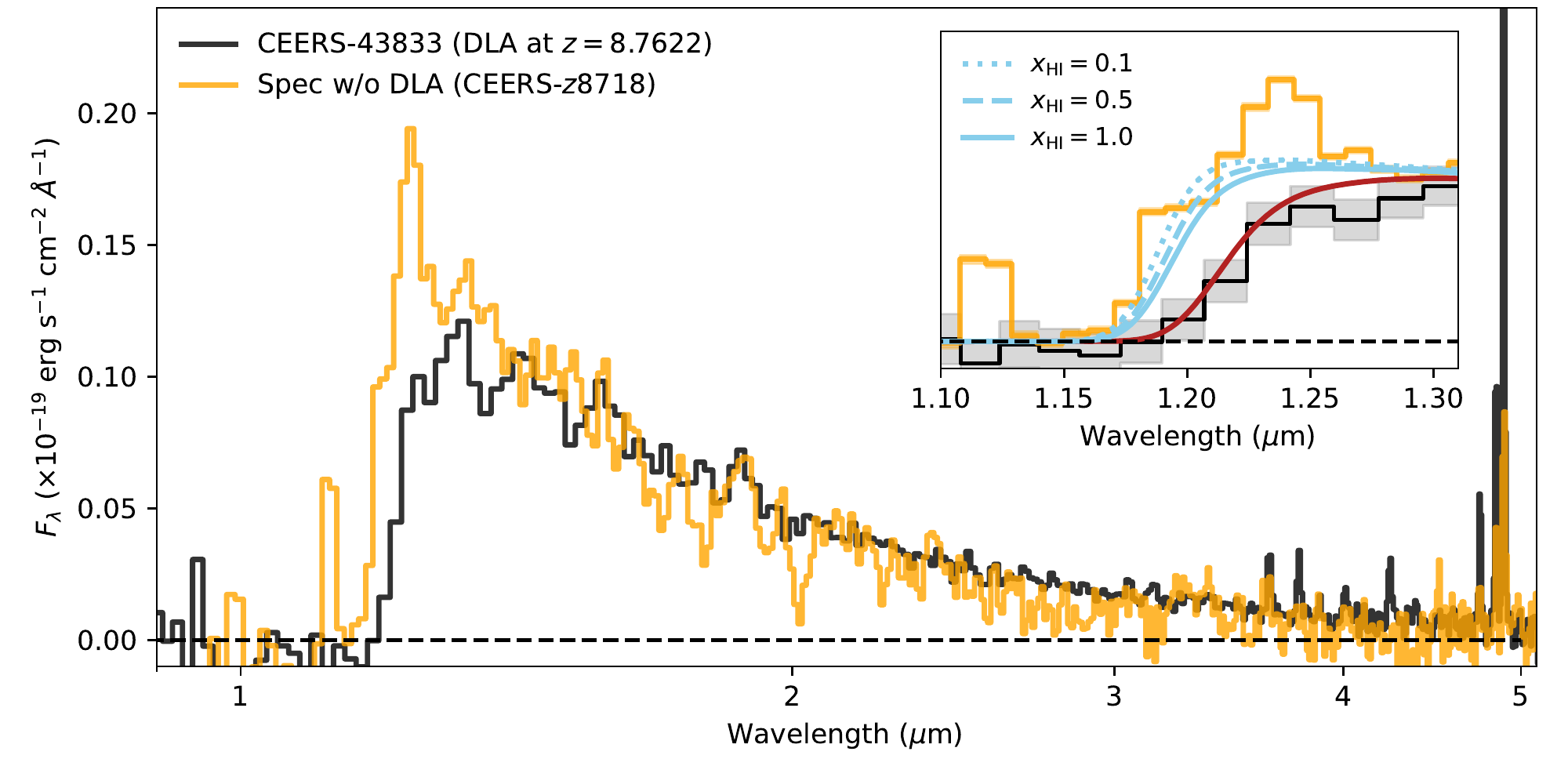}
    \includegraphics[width=0.49\textwidth]{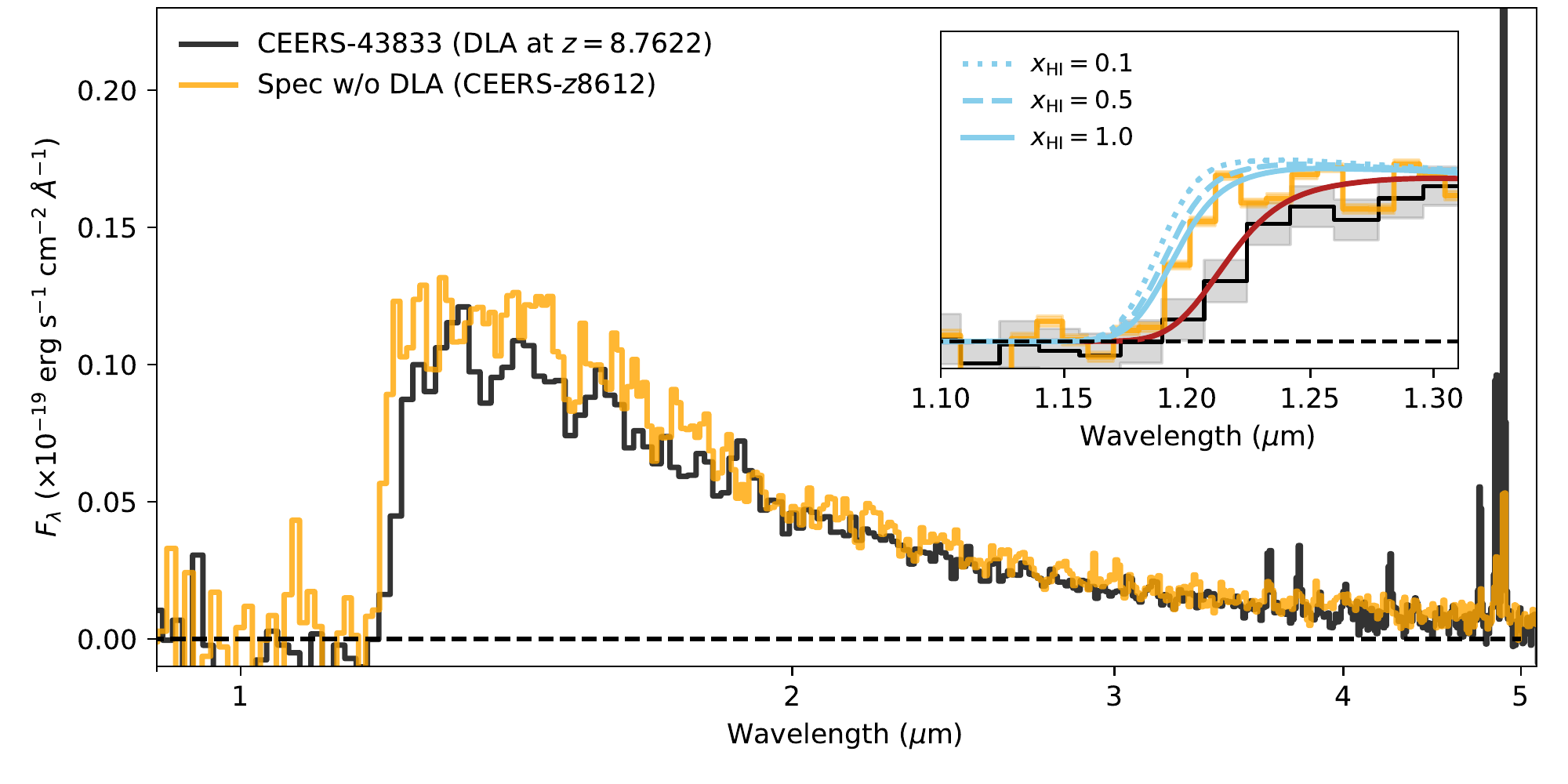}
    \includegraphics[width=0.49\textwidth]{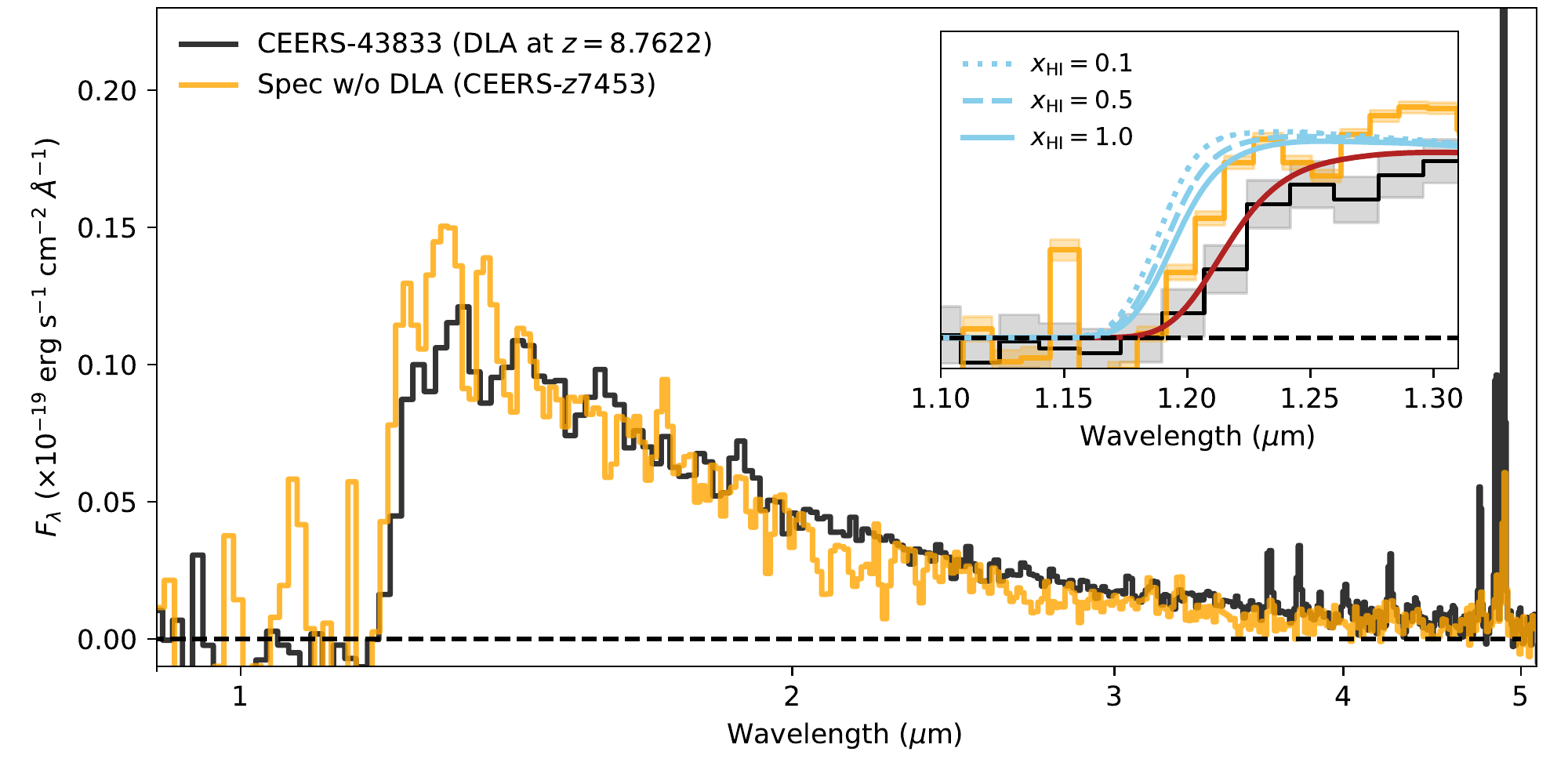}
    \label{fig:figs4}
\end{figure}

\noindent {\bf Fig. S4. \lya\ comparison plots of star-forming galaxies at $z\approx 8$.} CEERS-43833 (black) at $z=8.7622$ with a clear DLA feature is compared to a set of four galaxies from the literature \cite{Heintz22b} at similar redshifts and with equal or higher spectral quality. The IGM and DLA models are identical to Fig.~1, but here convolved with the spectral resolution of the JWST/NIRSpec prism observations. The models and comparable spectra exclude the broad \lya\ feature observed in CEERS-43833, and the two other galaxies at $z>9$, as being induced by instrumental broadening effects.

\clearpage

\begin{figure}
    \centering
    \includegraphics[width=0.8\textwidth]{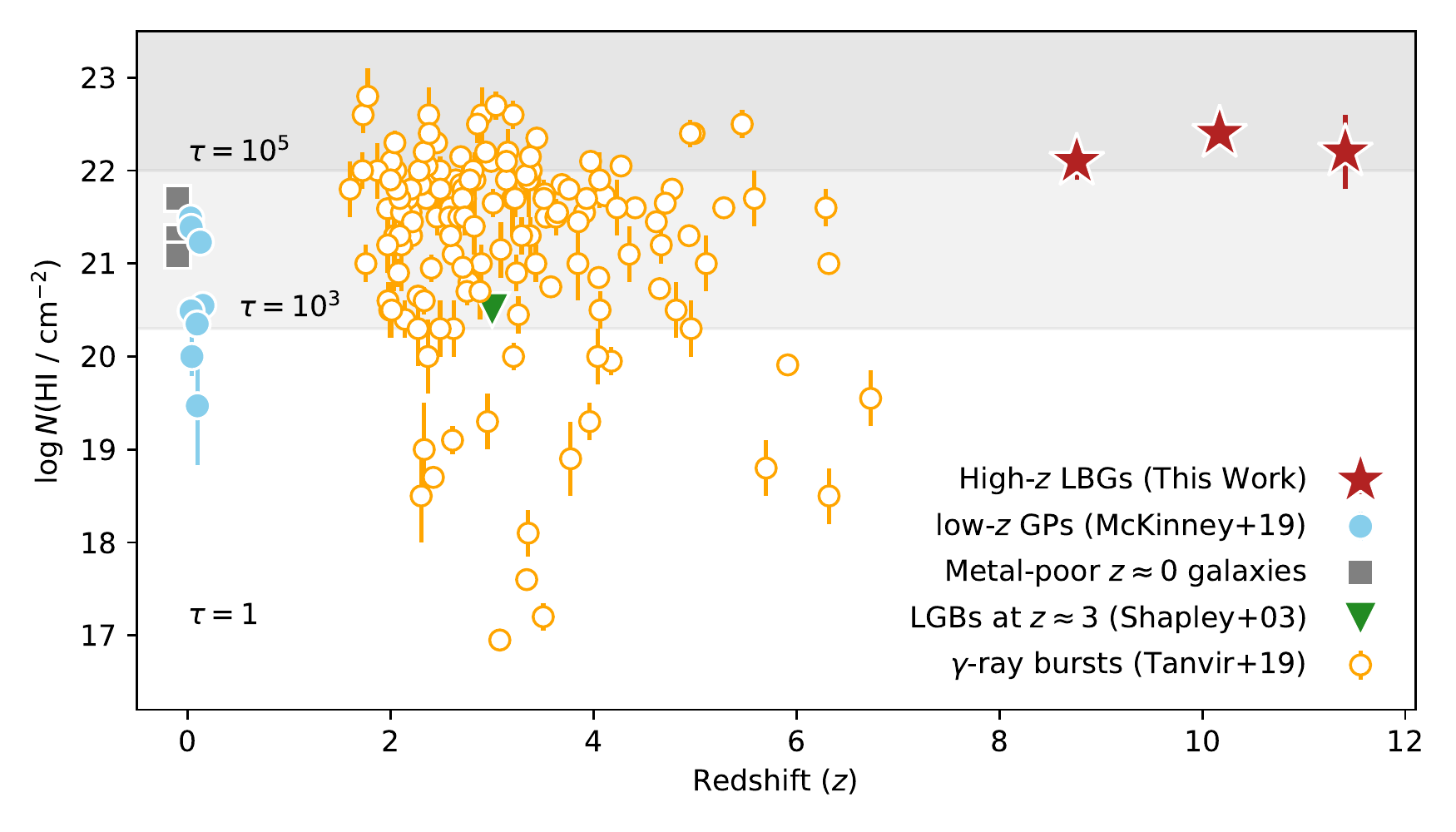}
%    \caption{}
    \label{fig:figs5}
\end{figure}

\noindent {\bf Fig. S5. \hi\ column density distribution as a function of redshift.} The symbol notation follow Fig.~2, but here we include additional constraints on the maximum $N_{\rm H\,\textsc{i}}$ observed for Lyman-break galaxies at $z\approx 3$ \cite{Shapley03} shown as the green triangle and the expanded set of $\gamma$-ray burst \hi\ absorbers at $z\sim 2-6$ \cite{Tanvir19} shown as the orange circles. The DLA demarcation region, $N_{\rm H\,\textsc{i}}>2\times 10^{20}\,$cm$^{-2}$, is marked by the light-grey area and the dark-grey region highlights the lower bound on $N_{\rm H\,\textsc{i}} > 10^{22}$\,cm$^{-2}$, corresponding to an optical depth of $\tau > 10^{5}$ at the Lyman limit, in the three $z=9-11$ galaxies analyzed in this work.

\clearpage

\begin{figure}
    \centering
    \includegraphics[width=0.8\textwidth]{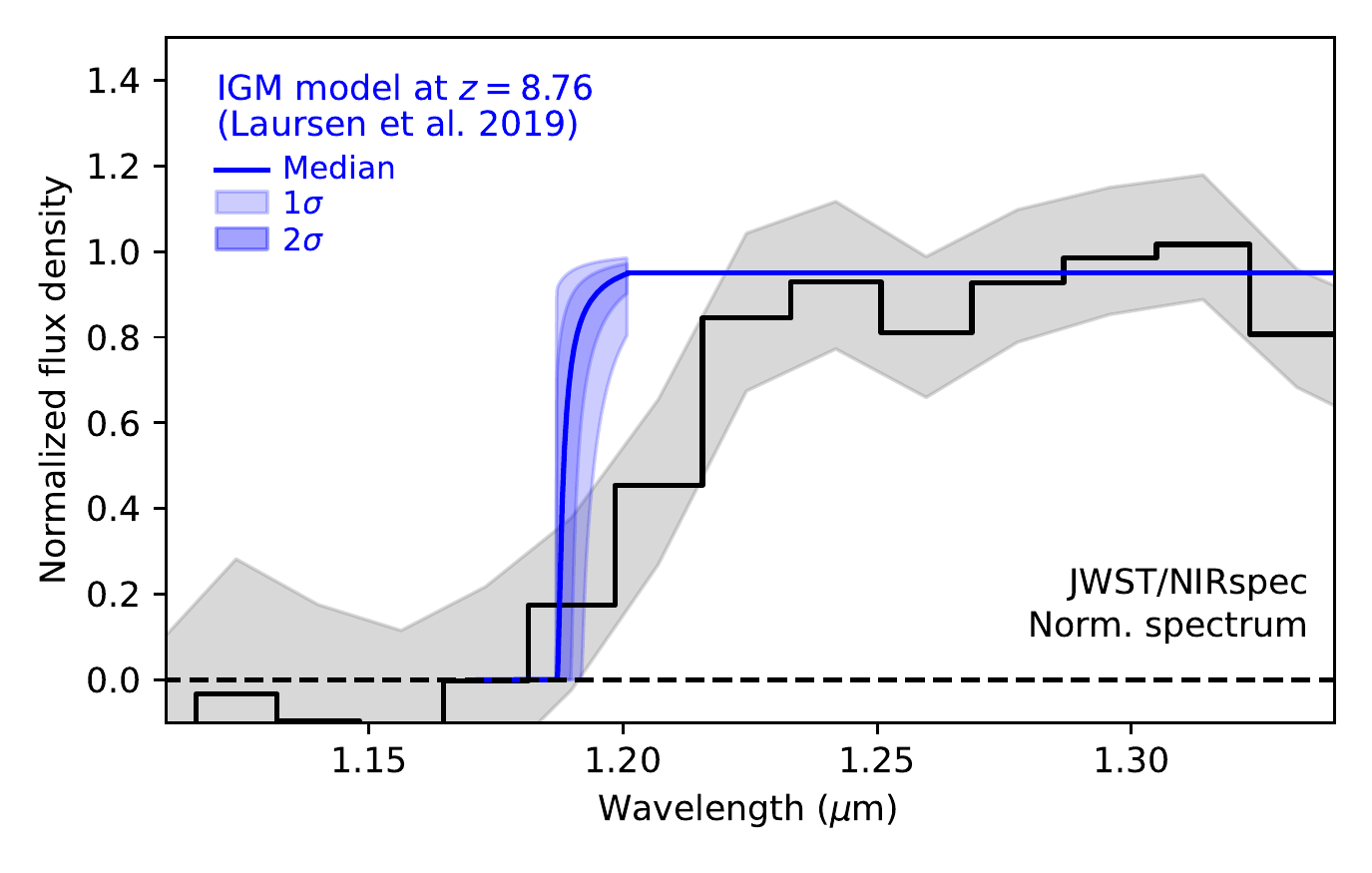}
%    \caption{}
    \label{fig:figs6}
\end{figure}

\noindent {\bf Fig. S6. IGM transmission curves from model galaxy sightlines.} The normalized spectrum of CEERS-43833 (black) and associated error spectrum (grey) is compared to the median (blue line) and $1\sigma$ and $2\sigma$ distributions of IGM transimissions curves modelled from typical galaxy sightlines at $z\simeq 8.8$, out to $10 r_\mathrm{vir}$, based on cosmological hydro-simulations and theoretical radiative transfer of ionizing UV and \lya\ photons \cite{Laursen19}. 

\end{document}